**"Scenario-Based Optimization of Network Resilience: Integrating Vulnerability Assessments and Traffic Flow"**

S. Saei[a], N. Tajik[b]

[a,b]Department of Industrial and System Engineering, Mississippi State University, Starkville, MS, United States [1]

Infrastructure networks face increasing risks from natural hazards and vulnerabilities in their design, making it crucial to develop methods for assessing and enhancing their resilience. This paper presents a scenario-based framework to evaluate network vulnerability by integrating local measures with topological analysis, allowing the assessment of each node's criticality in maintaining network integrity during disruptions. The framework compares structural properties with established network standards, identifying optimization opportunities to strengthen resilience. Traffic flow is incorporated using the Bureau of Public Roads (BPR) function to enhance the network's ability to withstand disruptions. A two-stage stochastic programming model is employed to capture uncertainties associated with vulnerability, ensuring robust network performance under diverse disruption scenarios. The approach balances risk-neutral (RN) and risk-averse (RA) strategies, with the hybrid risk-neutral and risk-averse (RNRA) model offering a comprehensive framework for managing network performance across various scenarios. Key factors such as node connectivity, disruption probability, and affected nodes are considered, highlighting the importance of fortifying critical nodes to prevent cascading failures. The proposed approach provides a robust method for enhancing network resilience by minimizing undelivered demand and optimizing overall performance under uncertain conditions.

[1] Corresponding Author Email: ss4646@msstate.edu

## 1. Introduction

From Hollnagel's perspective, the cornerstones of resilience involve several key aspects: "Knowing What to Do," (preparedness and response strategies), "Knowing What to Look For," (monitoring potential risks and disruptions), "Knowing What to Expect," (anticipating disruptions through vulnerability analysis), and "Knowing What Has Happened," (learning from past events to enhance future resilience). Collectively, these aspects contribute to a proactive and comprehensive approach to maintaining and strengthening system robustness and recovery capabilities (Hollnagel & others, 2011). In his book "Resilience Engineering in Practice: A Guidebook, (Hollnagel, 2013)" Hollnagel delves into practical applications of these principles in various contexts. Despite over two decades of application in transportation systems, there is still no widely agreed upon definition of system functionality for urban road transportation when it comes to resilience (Q.-L. Lu et al., 2024). Table 1 summarized definitions of resilience in transportation systems.

Table 1. Definition of resilience in transportation systems

| Definition | Area |
|---|---|
| The ability to prepare for changing conditions and withstand, respond to, and rapidly recover from disruptions (Scope, 2015) . | Resilience in transportation systems |
| Integrate resources from various service providers to manage disruptions effectively, ensuring minimal disruption impact and rapid recovery (Amghar et al., 2024) . | Resilience as a Service (RaaS) in transportation networks |
| The ability of an urban road transportation system to prepare for different kinds of disruptions, effectively serve vehicles, and recover rapidly to its optimal serving rate (i.e., trip completion rate) (Q.-L. Lu et al., 2024). | Traffic resilience |
| A property allowing the system to rebalance supply and demand functions in imbalance situations (J. Wang et al., 2013). | Networked service system |
| Resilience is contrary to efficiency, which means fewer routes and lower inventory, making systems less resilient. Efficient systems assume disruptions are rare, but climate change renders disruptions more frequent and unpredictable, necessitating a balance between efficiency and resilience (Rodrigue, 2024). | Transportation Systems |

Transportation resources are often prioritized for emergency and delivery vehicles and support freeway investments, contributing to car dependency and urban sprawl. This prioritization aligns with Transportation Demand Management (TDM) objectives by promoting efficient travel modes and applying pricing strategies to alleviate congestion, as exemplified by the Green Transportation Hierarchy, which favors non-motorized and high-occupancy vehicles. In fact, TDM is the art of influencing travel behavior to prevent the need for costly expansions of the transportation infrastructure (Litman, 2007). Such an approach influences policy and planning decisions, impacting the design and management of public roads and parking facilities to support sustainable transportation. Having a diverse array of transportation modes, such as walking, cycling, and transit, and maintaining redundant routes ensure service continuity if one component fails. The system must also be flexible and adaptable to changing conditions, with robust communication channels to provide timely information and coordinate responses during emergencies (Ferguson, 1990).

In an uncertain future, resilience strategies must prepare for diverse scenarios, including improbable events with serious consequences (Institute, 2019). Strengthening resilience requires identifying vulnerabilities—system weaknesses susceptible to disruptions from natural disasters, resource shortages, or malfunctions—to ensure readiness for both expected challenges and unforeseen events.

In the study of networks, variables can often be categorized into three pivotal domains: connectivity, accessibility, and criticality. In the domain of connectivity, the focus is on the direct links between nodes. These metrics generally describe how nodes are directly connected to each other and how influential they are within the network. Accessibility, on the other hand, deals with how easily nodes can be reached or accessed within the network. These metrics illuminate the efficiency and speed with which information or resources can traverse the network. In the domain of criticality, understanding the network's robustness and resilience, particularly when subjected to failures or disruptions, is crucial. Each of these 27 features plays a role in deciphering the intricate structure and behavior of networks, as shown in Table 2 (details of formula are in Appendix A).

Table 2. Local Measures Categories

| Connectivity | Accessibility | Criticality |
|---|---|---|
| Neighborhood Connectivity ($NC_i$) | Page Rank ($PR_i$) | Total Undelivered Demand after node j disrupted ($UD\varphi_i$) |
| Phi Node Centrality ($\varphi_i$) | Harmonic Centrality($HC_i$) | Group Centrality ($GrC_i$) |
| Eigenvector Centrality ($EI_i$) | Katz Centrality ($KC_i$) | Total Changes in the Path distance ($\Delta d_i^{OD}$) |
| Proportional Flow ($PF\varphi_i$) | Closeness Centrality ($CC_i$) | Average Path Distance ($AP_i$) |
| Degree Centrality ($DC_i$) | Exposure ($Expo_i$) | Average Path Distance after a node disruption ($APD_{\nexists i \in N}^{OD}$) |
| Indegree Centrality ($DC_i^+$) | Betweenness Centrality ($BC_i$) | Segmentwise ($seg_i$) |
| Outdegree Centrality ($DC_i^-$) | Aggregate Measure ($AG_i$) | Tsallis entropy-based redundancy ($TE_i$) |
| Weighted Node Measure after a node disruption ($GW_i$) | Average Rating ($AR_i$) | Star Tsallis entropy-based redundancy ($TE_i^*$) |
| Wighted Node ($W_i$) | | Complexity Measure Distribution ($CMT_i$) |
| | | Complexity Measure -Tsallis ($CM_i$) |

## 2. Literature Review

In the intricate world of transportation logistics, two-stage stochastic programming has become an essential tool for managing uncertainties and optimizing decision-making processes. Bureau of Public Roads (BPR) function plays a significant role in traffic modeling by describing the relationship between traffic flow and travel time, which is essential for transportation network planning and optimization. Gore et al., (2023) and Mattsson & Jenelius, (2015) note that the BPR function's ability to model travel time as a function of traffic flow provides critical insights into congestion dynamics and network performance. The BPR function's ability to account for travel time variability under different traffic conditions makes it a valuable tool for evaluating system-wide performance in transportation planning (Gore et al., 2023; Lien et al., 2016; N. Liu et al., 2013; Mattsson & Jenelius, 2015; Shen et al., 2022; Wong & Wong, 2016; J. Zhang et al., 2019). Recent advancements have expanded the applicability of the BPR function through the integration of stochastic elements, capturing the inherent variability in travel times due to factors such as accidents, weather conditions, and demand surges. The combination of these stochastic models with two-stage programming approaches offers a more nuanced understanding of transportation network behavior under uncertainty, leading to more resilient and adaptive planning strategies in the field of transportation logistics. A study by Lu et al., (2018) develops a mean-risk two-stage stochastic programming model to optimize the retrofitting and travel costs for transportation networks against extreme events like earthquakes, incorporating different risk preferences of decision-makers. The model uses Conditional Value-at-Risk (CVaR) for risk assessment and employs Generalized Benders Decomposition to tackle the nonconvexities of the optimization problem. Hu et al., (2021) introduces a sophisticated trilevel, two-stage, multi-objective stochastic model tailored for optimizing retrofit investments in road networks to mitigate damage from roadside tree blowdown during tropical cyclones. The model intricately uses the Bureau of Public Roads (BPR) function to estimate traffic flow and travel time dynamics within a stochastic framework, optimizing decisions across three hierarchical levels: strategic retrofit planning, repair sequencing, and traffic management, all streamlined by a random forest algorithm for computational efficiency. Lessan & Kim, (2022) examine stochastic optimization and Conditional Value-at-Risk (CVaR), the approach optimizes evacuation instructions across possible scenarios to address compliance uncertainty, demonstrating that purely risk-averse strategies may not always yield the most efficient outcomes in low-compliance scenarios. Zhuang et al., (2023) develops a robust two-stage stochastic programming model that enhances logistics networks' resilience against disruptions by optimizing decisions such as opening new lines and rerouting, while utilizing Conditional Value at Risk (CVaR) to mitigate operational risks. Lin et al., (2024) present a bilevel stochastic optimization model aimed at enhancing the resilience of transportation infrastructures by optimizing restoration schedules under uncertainty. The model incorporates Conditional Value at Risk with regret (CVaR-R) to minimize the risks associated with worst-case scenarios, providing a more robust framework for decision-makers to efficiently restore transportation networks following disruptive events.

## 3. Problem Description

In this problem, a decision-maker aims to optimize the design and protection of a transportation network under uncertain conditions, with the objective of minimizing the total expected cost, which encompasses travel time, penalties for unmet demand, and investments in fortification. The transportation network $G(V, E)$ is a mathematical model that $V$ is the set of nodes, where each node $i \in V$ represents a specific location, such as an intersection, station, or terminal within the transportation system. $E$ is the set of edges, where each edge $l \in E$ represents a direct connection or route between two nodes in the network, such as a road, railway, or flight path. Each edge $l \in E$ is associated with a capacity $V_l$ and a free-flow travel time. Under uncertainty, two types of decision variables are considered: the first stage and the second stage. The first-stage variables as fortification are decisions made before the uncertainty is revealed. The fortification of the network can mitigate these disruptions, reducing the disruption rate on the affected links, with the level of reduction depending on the scenario and the specific links fortified. The cost of fortification varies across the network, influencing decision-making on how to best allocate resources to maintain network resilience under various disruption scenarios. In each scenario, different links in the network experience disruptions, with disruption rates varying for each link. The second-stage variables, decisions are determined after the uncertainty is realized. The model uses a finite set of scenarios to define which nodes in the network are affected and to what extent their capacities as disruption rate are reduced. In fact, the impact of these disruptions is modeled for specific nodes and their associated incoming/outgoing links, providing a detailed understanding of potential network vulnerabilities.

### 3.1 Model Formulation Without disruption (business-as-usual)

The sets and indexes, parameters, and decision variables are in table 4:

Table 3. The Sets and Indexes, parameters, and decision variables

| **Sets and Indexes** | |
|---|---|
| $\mathcal{I}$ | Set of nodes (indexed by $i, j$ ) |
| $\mathcal{L}$ | Set of links |
| $\mathcal{O} \subseteq \mathcal{I}$ | Set of origin nodes [subset of all nodes as origins] |
| $\mathcal{D} \subseteq \mathcal{I}$ | Set of destination nodes |
| $\mathcal{R}^{o \to d}$ | Set of possible routes between origin $o \in \mathcal{O}$ and destination $d \in \mathcal{D}$ (indexed $R$) |
| $\mathcal{K}_r^{o \to d} \subseteq \mathcal{L}$ | Subset of the links belonging to route $r$ associated with origin $o \in \mathcal{O}$ and destination $d \in \mathcal{D}$ |

| **Parameters** | |
|---|---|
| $T_l^0$ | Travel time in free-flow condition of link $l \in \mathcal{L}$ |
| $V_l$ | Capacity of link $l \in \mathcal{L}$ |
| $\alpha, \beta$ | Coefficients of congestion function, where α influences the flow's impact on travel time and β controls the non-linearity of the congestion effect. |
| $d^{o \to d}$ | Delivered demand between origin $o \in \mathcal{O}$ and destination $d \in \mathcal{D}$ |

| Decision Variables | |
|---|---|
| $f_r^{o \to d}$ | Flow between origin $o \in \mathcal{O}$ and destination $d \in \mathcal{D}$ via route $r \in \mathcal{R}^{o \to d}$ |
| $T_r^{o \to d}$ | Total travel time between origin $o \in \mathcal{O}$ and destination $d \in \mathcal{D}$ via route $r \in \mathcal{R}^{o \to d}$ |
| $h_l$ | Traffic on link $l \in \mathcal{L}$ |
| $t_l$ | Travel time on link $l \in \mathcal{L}$ |
| $z$ | Maximum travel time |
| $u^{o \to d}$ | Undelivered demand |

Objective function:

$$Min\ \omega_1 \frac{z}{max\,(t_l)} + \omega_2 \frac{1}{max(\sum_{o \in \mathcal{O}} \sum_{d \in \mathcal{D}} d^{o \to d})} \sum_{o \in \mathcal{O}} \sum_{d \in \mathcal{D}} u^{o \to d}$$

Subject to:

$$\sum_{r \in \mathcal{R}^{o \to d}} f_r^{o \to d} \le d^{o \to d} \qquad \forall (o \in \mathcal{O}, d \in \mathcal{D}) \qquad (1)$$

$$h_l = \sum_{o \in \mathcal{O}} \sum_{d \in \mathcal{D}} \sum_{r \in \mathcal{R}^{o \to d}:\, l \in \mathcal{K}_r^{o \to d}} f_r^{o \to d} \qquad \forall (l \in \mathcal{L}) \qquad (2)$$

$$t_l = T_l^0 \left(1 + \alpha \left(\frac{h_l}{V_l}\right)^{\beta}\right) \qquad \forall (l \in \mathcal{L}) \tag{3}$$

$$T_r^{o \to d} = \sum_{l \in \mathcal{K}_r^{o \to d}} t_l \qquad \forall (o \in \mathcal{O}, d \in \mathcal{D}, r \in \mathcal{R}^{o \to d}) \tag{4}$$

$$z \geq T_r^{o \to d} \qquad \forall (o \in \mathcal{O}, d \in \mathcal{D}, r \in \mathcal{R}^{o \to d}) \tag{5}$$

$$f_r^{o \to d} \geq 0 \tag{6}$$

$$h_l \geq 0$$
$$t_l \geq 0$$
$$T_r^{o \to d} \geq 0$$
$$z \geq 0$$

This model shows a network traffic optimization problem. The objective function minimizes the maximum travel times and maximizes (or minus minimum) delivered demand from origin to destination. This reflects a common transportation network flow optimization challenge, with a tradeoff between travel times and undelivered demand (the less travel time, the more undelivered demand). Constraint 1 ensures that flow guarantees delivered demand between origin $o \in \mathcal{O}$ and destination $d \in \mathcal{D}$. Constraint 2 ensures that the traffic of the link is based on the sum of flow for all links belonging to a route associated with origin $o \in \mathcal{O}$ and destination $d \in \mathcal{D}$, for all origin-destination pairs. In other words, if the link is not present in a route, it is not necessary to consider that route since the traffic does not pass that link. Constraint 3 calculates the travel time on each link $l$, which is the function of traffic called as the Bureau of Public Roads (BPR) congestion parameters $\alpha$ and $\beta$. Usually, $\alpha$=0.15 and $\beta$=4 to consider the sensitivity of the traffic on link $l$ and capacity. Constraint 4 determines the total travel time for all links belonging to a route associated with origin $o \in \mathcal{O}$ and destination $d \in \mathcal{D}$. Constraints 5 ensures the maximum travel time of Total travel for all links belonging to a route associated with origin $o \in \mathcal{O}$ and destination $d \in \mathcal{D}$. Constraint 6 guarantees that all decision variables in the model, including travel times, flow, and delivered demand, remain non-negative, reflecting realistic physical conditions.

### 3.2 Proposed Model to Cope with Disruption

In modern infrastructure networks, disruptions can lead to significant inefficiencies and failures, impacting both service delivery and system resilience. Addressing these challenges requires the development of robust optimization models that can effectively mitigate the consequences of such disruptions. In this study, we present a model designed to minimize the adverse effects of network disruptions by optimizing critical parameters such as undelivered demand and maximum travel time. By considering a wide range of disruption scenarios, the model incorporates the stochastic nature of real-world events, allowing for strategic decisions on node fortification and resource allocation. Probabilistic weighting of scenarios ensures that both high-risk and low-risk disruptions are accounted for, providing a balanced and comprehensive approach to network resilience. In this section, we outline the two-stage stochastic programming approach based on both risk-neutral (RN) and risk-averse (RA) measures.

Assumptions:

- First stage decisions as “Here and now” are made before realization of the disruption based on the deterministic normal situation.
- Second decisions as “Wait and see” are made after realization of the uncertainty about the disaster scenario is revealed.
- Some nodes and associated links might be disrupted and become unavailable during and shortly after the disaster.

The fortification of the network can mitigate these disruptions, reducing the disruption rate on the affected links, with the level of reduction depending on the scenario and the specific links fortified. The cost of fortification varies across the network, influencing decision-making on how to best allocate resources to maintain network resilience under various disruption scenarios. In each scenario, different links in the network experience disruptions, with disruption rates varying for each link. For example, under scenario $\xi 1$, certain links experience up to 70% reduction in capacity. To counter these disruptions, we consider the option of fortifying the network, where fortification reduces the disruption rate at a cost, which varies for different parts of the network. This model is solved using Gourbi's mixed-integer programming solver, which can handle the complex structure of the TSSP problem. This implementation demonstrates how the theoretical TSSP framework can be applied to practical problems like network traffic distribution under uncertainty. Table 4 shows the sets and indexes, parameters, and decision variables.

**Notations:**

Table 4. The Sets and Indexes, parameters, and decision variables under uncertainty

| **Sets and Indexes** | |
|---|---|
| $\mathcal{L}_i$ | Subset of links connected to node $i \in \mathcal{I}$ |
| $\Xi$ | Set of Disruption Scenarios |
| $\mathcal{I}_\xi \subseteq \mathcal{I}$ | Subset of nodes affected in disruption scenario $\xi \in \Xi$ |
| **Parameters** | |
| $c_i$ | Cost of fortifying node $i \in \mathcal{I}$ |
| $\vartheta_{i\xi}$ | Percentage of disruption on node $i \in \mathcal{I}_\xi$ (links $l \in \mathcal{L}_i$) under scenario $\xi \in \Xi$ (in the case of having no fortification, the impact of disruption is $(1-\gamma_{l\xi})\vartheta_{l\xi}$) |
| $\gamma_{l\xi}$ | Percentage of decreasing disruption on links $l \in \mathcal{L}_i$ under scenario $\xi \in \Xi$ for $i \in \mathcal{I}_\xi$ (in the case of having fortification, the impact of disruption is $\vartheta_{l\xi}$) |
| $\pi_\xi$ | Chance/Probability of scenario $\xi \in \Xi$ |
| $M$ | A big enough number |
| $N_f$ | Minimum required number of fortifications |
| **Decision Variables** | |
| **First Stage (Here and Now)** | |
| $x_i$ | 1, if node $i \in \mathcal{I}$ is fortified; 0, else. |
| **Second Stage (Wait and See)** | |
| $f_{r\xi}^{o\to d}$ | Flow |
| $T_{r\xi}^{o\to d}$ | Total travel time between origin $o \in \mathcal{O}$ and destination $d \in \mathcal{D}$ via route $r \in \mathcal{R}^{o\to d}$ |
| $h_{l\xi}$ | Traffic on link $l \in \mathcal{L}$ |
| $t_{l\xi}$ | Travel time on link $l \in \mathcal{L}$ |
| $z_\xi$ | Maximum travel time under scenario $\xi \in \Xi$ |
| $u_\xi^{o\to d}$ | Undelivered demand under scenario $\xi \in \Xi$ |
| $V'_{l\xi}$ | Reminded capacity of link $l \in \mathcal{L}$ under scenario $\xi \in \Xi$ |

Objective Function:

$$\boldsymbol{MIN} \sum_{\xi \in \Xi} \pi_\xi \left( \omega_1 \frac{z_\xi}{t_{l\xi}} + \omega_2 \frac{1}{max(\sum_{o \in \mathcal{O}} \sum_{d \in \mathcal{D}} d^{o\to d})} \sum_{o \in \mathcal{O}} \sum_{d \in \mathcal{D}} u_\xi^{o\to d} \right) + \omega_3 \frac{1}{max_i(c_i)} \sum_{i \in \mathcal{I}} c_i x_i$$

Subject to:

$$\sum_{r \in \mathcal{R}^{o\to d}} f_{r\xi}^{o\to d} = d^{o\to d} - u_\xi^{o\to d} \qquad \forall (o \in \mathcal{O}, d \in \mathcal{D}, \xi) \qquad (7)$$

$$h_{l\xi} = \sum_{o \in \mathcal{O}} \sum_{d \in \mathcal{D}} \sum_{r \in \mathcal{R}^{o\to d}:\, l \in \mathcal{K}_r^{o\to d}} f_{r\xi}^{o\to d} \qquad \forall (l \in \mathcal{L}, \xi) \qquad (8)$$

$$t_{l\xi} \geq T_l^0 \left( 1 + \alpha \frac{h_{l\xi}^\beta}{V_l^\beta} \right) \qquad \forall (\xi \in \Xi, i \notin \mathcal{I}_\xi, l \in \mathcal{L}_i) \qquad (9)$$

$$t_{l\xi} \geq T_l^0 \left( 1 + \alpha \frac{h_{l\xi}^\beta}{\left( V_l \left( 1 - (1 - \gamma_{l\xi}) \vartheta_{l\xi} \right) \right)^\beta} \right) - (1 - x_i) M \qquad \forall (\xi \in \Xi, i \in \mathcal{I}_\xi, l \in \mathcal{L}_i) \qquad (10)$$

$$t_{l\xi} \geq T_l^0 \left( 1 + \alpha \frac{h_{l\xi}^\beta}{\left( V_l (1 - \vartheta_{l\xi}) \right)^\beta} \right) - x_i M \qquad \forall (\xi \in \Xi, i \in \mathcal{I}_\xi, l \in \mathcal{L}_i) \qquad (11)$$

$$T_{r\xi}^{o\to d} \geq \sum_{l \in \mathcal{K}_r^{o\to d}} t_{l\xi} \qquad \forall (o \in \mathcal{O}, d \in \mathcal{D}, r \in \mathcal{R}^{o\to d}, \xi) \qquad (12)$$

$$z_\xi \geq T_{r\xi}^{o\to d} \qquad \forall (o \in \mathcal{O}, d \in \mathcal{D}, r \in \mathcal{R}^{o\to d}, \xi) \qquad (13)$$

$$\sum_{i \in \mathcal{I}} x_i \leq N_f \qquad (14)$$

$$x_i \in \{0,1\} \qquad (15)$$
$$f_{r\xi}^{o\to d} \geq 0$$
$$h_{l\xi} \geq 0$$
$$t_{l\xi} \geq 0$$
$$T_{r\xi}^{o\to d} \geq 0$$
$$z_\xi \geq 0$$
$$V'_{l\xi} \geq 0$$

The optimization model enables a comprehensive analysis of network resilience and performance under various scenarios, considering factors such as link capacity, fortification decisions, and scenario-specific disruptions. The model aims to optimize the trade-off between network resilience investments and operational performance, providing valuable insights for strategic decision-making in complex network systems. The objective function minimizes a weighted sum of undelivered demand, maximum travel time, and fortification costs across all scenarios, with each scenario weighted by its probability. Constraint 7 ensures that flow guarantees delivered demand between origin $o \in \mathcal{O}$ and destination $d \in \mathcal{D}$ under disruption in each scenario. Constraint 8 calculates the traffic of the link as the sum of flows for all links belonging to a route associated with disrupted in each scenario. Constraints 9, 10, and 11 describe the Travel Time based on the Bureau of Public Roads (BPR) function. These constraints are linearized using a **Piecewise Linear Approximation (PLA)** approach for more accurate traffic flow modeling. Constraint 12 calculates the total travel time for each route. Constraint 13 determines the maximum travel time across all routes. Constraint 14 ensures a minimum number of fortifications, and Constraint 15 ensures all variables are non-negative.

### 3.3 Basic Two-Stage Stochastic Programming (TSSP) Problem:

Decision-making under uncertainty involves different attitudes towards risk. The basic TSSP model minimizes the overall cost by making an initial decision $x$ and then optimizing the second-stage decision $w$ based on the realization of uncertain parameters $\theta$. The risk-neutral TSSP approach considers the expectation of the second-stage cost across all possible realizations of uncertainty, effectively aiming to minimize the expected total cost without explicitly addressing risk. The scenario-based TSSP introduces a discrete set of scenarios, each with an associated probability, allowing for a more detailed analysis of various outcomes, thereby optimizing the expected cost over all scenarios. Model 16 represents a two-stage stochastic programming problem where the decision-maker first chooses $x$ as the first-stage decision variables, then $w$ is determined based on the realization of uncertain parameters $\theta = [g, D, h]$. The second stage captures the cost associated with the decisions under uncertainty.

$$\min_{x} fx + R(x;\boldsymbol{\theta}) \tag{16}$$

$$subject\ to{:}\ \ Kx \leq b$$

$$x \geq 0$$

in which:

$$R(x;\boldsymbol{\theta}) = \min_{w} gw$$

$$s.t.\ Aw + Dx \geq h$$

$$w \geq 0$$

Risk-Neutral TSSP models incorporate constraints to ensure feasibility under uncertainty. The probability distribution approach minimizes the sum of the initial decision cost and the expected second-stage cost, with the expectation taken over a distribution of uncertain parameters. The scenario-based approach assumes a finite number of scenarios ($\Xi$) for the uncertain parameters, each with a specific probability $\pi_{\xi}$. In the wait-and-see approach (Birge & Louveaux, 2011), each scenario is treated individually to find the optimal solution, as if random outcomes were known in advance. This approach generates a set of scenario-specific solutions, which are often combined into a single implementable solution using heuristic rules.

### 3.4 CVaR for Risk-Averse RA TSSP Decision Making:

Conditional Value-at-Risk (CVaR) offers a coherent and more stable alternative for quantifying tail risks that Value-at-Risk (VaR) may not fully capture. CVaR can assess the severity of losses beyond a predetermined threshold, providing a more conservative evaluation of potential financial losses (Rockafellar & Uryasev, 2002).. CVaR and its minimization formula developed by Rockafellar et al., (2000) enables risk-averse decision-making by explicitly incorporating the potential for extreme costs into the optimization problem, thus protecting the decision-maker against adverse outcomes. Model 17 introduces CVaR as a measure for risk aversion, targeting the minimization of expected costs in worst-case scenarios. This approach is particularly valuable in situations ensuring robustness against extreme outcomes. To make the problem more tractable, the CVaR term, the non-linear CVaR expression $\left[g_\xi w_\xi - \nu\right]^+$ is linearized.

$$\min_{x,w,\nu} fx + \overbrace{\nu + \frac{1}{\epsilon}\sum_{\xi \in \Xi} \pi_\xi \left[g_\xi w_\xi - \nu\right]^+}^{\mathrm{CVaR}_\epsilon[R]} \tag{17}$$

$$s.t.\ Kx \leq b$$

$$Aw_\xi + D_\xi x \geq h_\xi$$

The non-linear term, representing the excess cost beyond threshold $\nu$, is simplified with auxiliary variables $\tau_\xi$ and additional linear constraints, as shown in model 18 (Schultz & Tiedemann, 2006).

$$\min_{x,w,\nu} fx + \nu + \frac{1}{\epsilon}\sum_{\xi \in \Xi} \pi_\xi \tau_\xi \tag{18}$$

$$s.t.\ Kx \leq b$$

$$Aw_\xi + D_\xi x \geq h_\xi$$

$$\tau_\xi \geq g_\xi w_\xi - \nu$$

$$\tau_\xi \geq 0$$

### 3.5 RNRA-TSSP measure, Conditional Value at Risk (CVaR-based):

The CVaR-based RNRA-TSSP (Risk-Neutral/Risk-Averse TSSP) model extends this framework by introducing a parameter $\delta$, balancing the focus between minimizing expected costs and accounting for worst-case scenarios, providing a comprehensive and flexible approach to decision-making under uncertainty (figure 1).

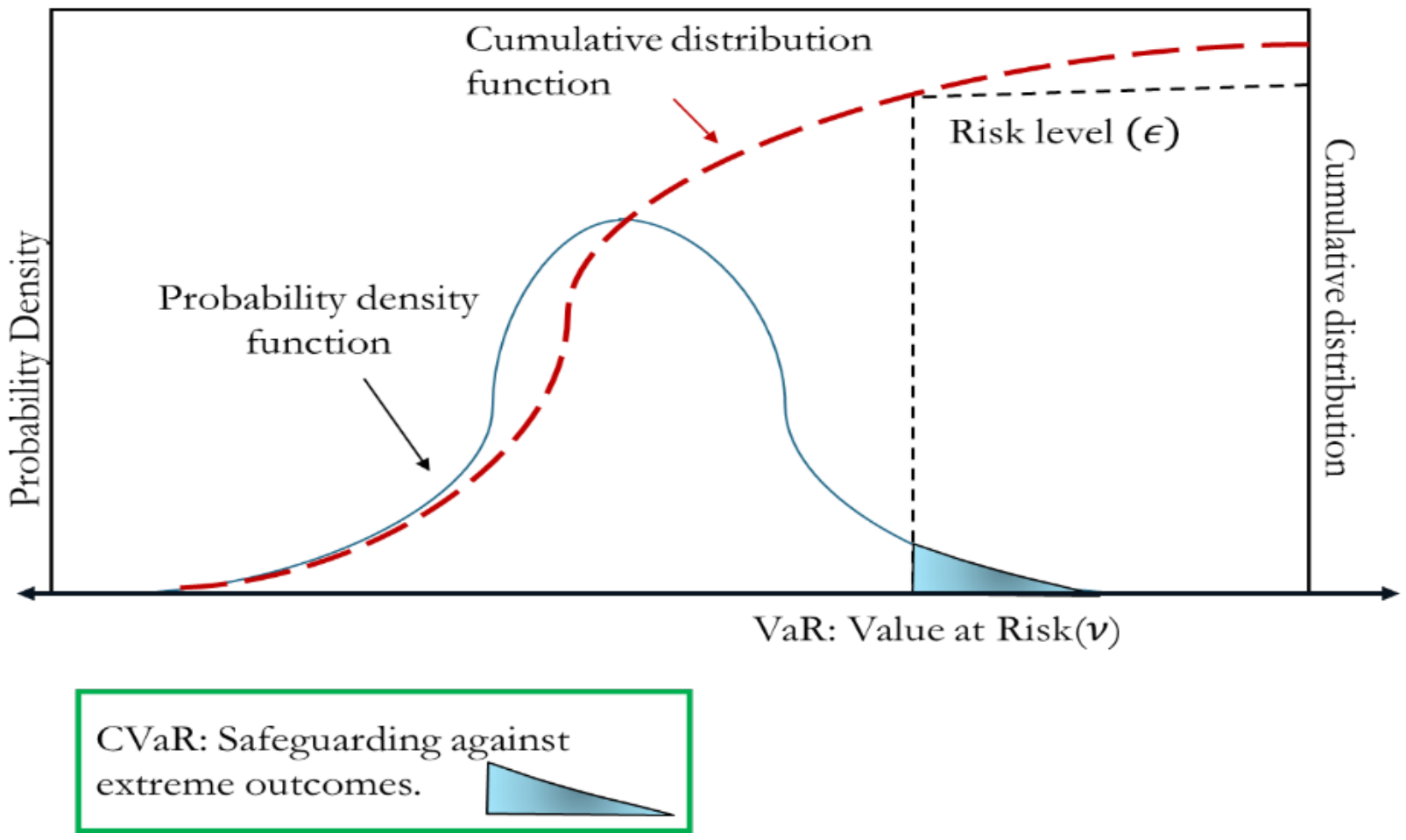

Figure 1. Concept of CvaR in TSSP

Constraint 19 ensures that the model is robust against extreme risks while maintaining tractability, making it a powerful tool for complex stochastic optimization problems.

$$\min_{x,w,v} fx + (1-\delta)\sum_{\xi\in\Xi}\pi_\xi\, g_\xi w_\xi + \delta\,(\nu + \frac{1}{\epsilon}\sum_{\xi\in\Xi}\pi_\xi\tau_\xi) \tag{19}$$

$$s.t.\ Kx \leq b$$

$$Aw_\xi + D_\xi x \geq h_\xi$$

$$\tau_\xi \geq g_\xi w_\xi - \nu$$

$$\tau_\xi \geq 0$$

$$0 \leq \delta \leq 1$$

Where $\delta$ controls the balance between risk-neutral ($\delta = 0$) and risk-averse ($\delta = 1$) decision-making.

- Risk-Neutral ($\delta = 0$): The model minimizes the expected costs without considering the potential for extreme losses, focusing entirely on average outcomes.

- Risk-Averse ($\delta$ = 1): The model emphasizes minimizing the impact of worst-case scenarios, effectively safeguarding against extreme outcomes by incorporating CVaR.
- Balanced Approach ($0 < \delta < 1$): The model strikes a balance between minimizing expected costs and protecting against extreme losses, providing a more nuanced approach to decision-making under uncertainty.

### 3.5.1 RA-TSSP measure, Conditional Value at Risk (CVaR-based) for proposed model:

The RA-TSSP (Risk-Averse Two-Stage Stochastic Programming) measure using CVaR focuses on minimizing the expected risk associated with extreme scenarios. The proposed model represented in Equation 20-23 is as follows:

$$\boldsymbol{MIN}\ \nu + \frac{1}{\epsilon}\sum_{\xi \in \Xi} \pi_\xi \tau_\xi + \omega_3 \frac{1}{max_i(c_i)} \sum_{i \in \mathcal{I}} c_i x_i \tag{20}$$

$$\tau_\xi \geq \omega_1 \frac{z_\xi}{max(T_r^{o \to d})} + \omega_2 \frac{1}{max(\sum_{o \in \mathcal{O}} \sum_{d \in \mathcal{D}} d^{o \to d})} \sum_{o \in \mathcal{O}} \sum_{d \in \mathcal{D}} u_\xi^{o \to d} - \nu \tag{21}$$

$$\tau_\xi \geq 0 \qquad \nu \geq 0 \tag{22}$$

$$+\ \text{Eqs(7)-(15)} \tag{23}$$

Where:

- $\nu$ is a decision variable that represents the baseline risk.
- $\epsilon$ is the risk-aversion parameter, which controls how much risk the decision-maker is willing to accept.
- $\pi_\xi$ represents the probability of each scenario $\xi \in \Xi$.
- $\tau_\xi$ captures the excess risk in each scenario beyond the baseline.
- $\omega_3$ and $c_i$ are cost parameters associated with decision variables $x_i$.

The constraints governing the excess risk ($\tau_\xi$) are given by Equations 20, 21. Equation 22 indicates that the formulation is supplemented by additional constraints from Eqs. (7) -(15), which handles operational decisions.

### 3.5.2 RARN-TSSP measure for proposed model:

The RARN-TSSP (Risk-Averse and Risk-Neutral Two-Stage Stochastic Programming) measure combines both risk-averse and risk-neutral components in a single framework, which allows balancing between risk minimization and cost efficiency. The objective function in Equation 24 is:

$$\begin{aligned} \boldsymbol{MIN}\ \ \omega_3 \frac{1}{max_i(c_i)} \sum_{i \in \mathcal{I}} c_i x_i \\ + (\mathbf{1} - \boldsymbol{\delta}) \sum_{\xi \in \Xi} \pi_\xi \left[ \omega_1 \frac{z_\xi}{max(T_r^{o \to d})} \right. \\ \left. + \omega_2 \frac{1}{max(\sum_{o \in \mathcal{O}} \sum_{d \in \mathcal{D}} d^{o \to d})} \sum_{o \in \mathcal{O}} \sum_{d \in \mathcal{D}} u_\xi^{o \to d} \right] \\ + \boldsymbol{\delta} \left( \nu + \frac{1}{\epsilon} \sum_{\xi \in \Xi} \pi_\xi \tau_\xi \right) \end{aligned} \tag{24}$$

$$\tau_\xi \geq \omega_1 \frac{z_\xi}{max(T_r^{o \to d})} + \omega_2 \frac{1}{max(\sum_{o \in \mathcal{O}} \sum_{d \in \mathcal{D}} d^{o \to d})} \sum_{o \in \mathcal{O}} \sum_{d \in \mathcal{D}} u_\xi^{o \to d} - \nu \tag{25}$$

$$\tau_\xi \geq 0 \tag{26}$$

$$\nu \geq 0$$

$$+\text{Eqs(7)-(15)} \tag{27}$$

Where:

- $\delta$ is a risk-splitting parameter that balances the risk-neutral and risk-averse components.
- The first term corresponds to cost minimization.
- The second term incorporates the risk-neutral part, while the third term represents the CVaR-based risk-averse component (similar to the RA-TSSP measure).

Similar to the RA-TSSP, this formulation also includes additional constraints from Eqs. (7) -(15).

## 4. Solution Approach:

In Two-Stage Stochastic Programming, uncertainty can be addressed by either using a probability distribution function (PDF) or a scenario-based approach. The PDF method represents uncertainties as a continuous distribution, capturing a wide range of potential outcomes with associated probabilities, ideal for analyzing expected system behavior when data on uncertainty distributions is available. Alternatively, the scenario-based approach defines a finite set of possible scenarios, each with a specific probability, to represent distinct conditions from typical to extreme cases. Figure 2 illustrates how these approaches apply to Risk-Neutral, Risk-Averse, and Hybrid (Risk-Neutral + Risk-Averse) TSSP models: the Risk-Neutral model minimizes average expected costs, the Risk-Averse model mitigates worst-case losses, and the Hybrid model strikes a balance, optimizing for average outcomes while also protecting against extreme risks. Together, these methods provide a robust framework for understanding and managing network performance under uncertainty.

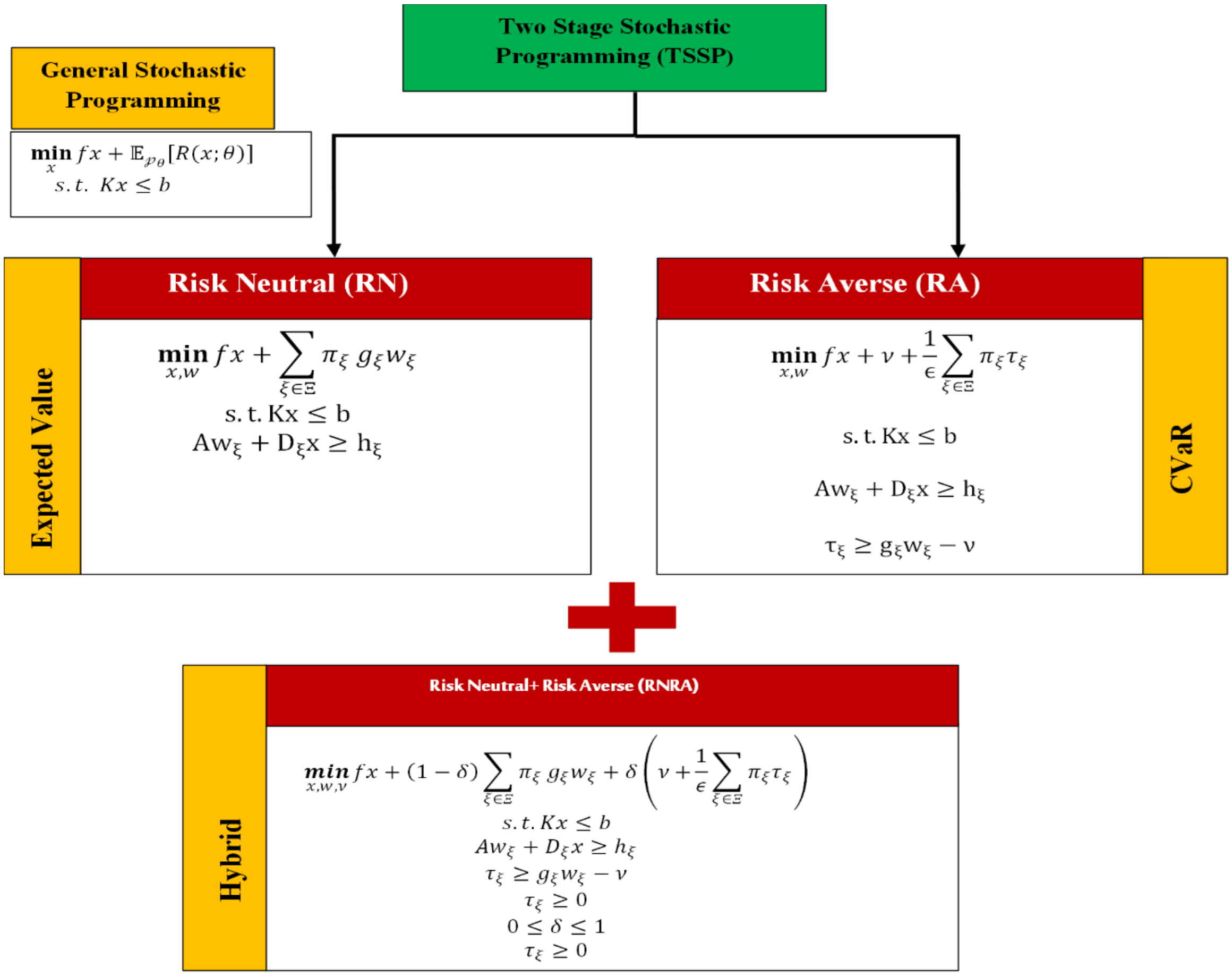

Figure 2. Risk-Neutral, Risk Averse, and Risk Neutral+ Risk Averse TSSP Models

We utilized the Gurobi solver 11.0.3 to solve a mixed-integer programming (MIP) model, leveraging advanced optimization techniques, such as cutting planes, to refine the problem formulation. By running multiple iterations with different delta values, we gained valuable insights into the influence of these changes on the solution, uncovering critical patterns and behaviors. This iterative exploration helped guide parameter tuning, leading to enhanced model performance and efficiency.

### 4.1. Piecewise Linear Approximation (PLA)

The Piecewise Linear Approximation (PLA) method is used to linearize single-variable nonlinear terms, such as $f(x)$ where $L \leq N \leq U$ (Dantzig, 2016). PLA effectively transforms nonlinear functions into linear approximations, making them suitable for linear programming solvers. This approach is particularly useful for convex functions in maximization problems or concave functions in minimization problems, since it guarantees global optimality in these cases. However, the accuracy of the approximation depends on the number of breakpoints used. Increasing N improves accuracy but also increases computational complexity due to the additional variables and constraints. Therefore, there is a trade-off between approximation accuracy and problem size that must be considered when implementing PLA. The process for PLA is as follows:

Phase 1: Determine N and create $\mathrm{N}+1$ points in the interval $[L, U]$:

$$\mathrm{a_i} = \mathrm{L} + \frac{\mathrm{i}}{\mathrm{N}}(\mathrm{U} - \mathrm{L}) \quad \mathrm{i} = 0,1,2, \dots \mathrm{N} \tag{28}$$

Phase 2: Calculate the function value for each of the N+1 points from Phase 1:

$$\mathrm{b_i} = \mathrm{f(a_i)} = \mathrm{f}\left(\mathrm{L} + \frac{\mathrm{i}}{\mathrm{N}}(\mathrm{U} - \mathrm{L})\right) \quad \mathrm{i} = 0,1,2, \dots \mathrm{N} \tag{29}$$

Phase 3: Define N+1 positive variables $\lambda_i \quad i = 0,1,2, \dots N$.

Phase 4: Express $x$ as a convex combination of $a_i$: $x = \sum_{i=0}^{N} \lambda_i a_i$ and $\sum_{i=0}^{N} \lambda_i = 1$Phase 5: Approximate the function range as a convex combination of $b_i = f(a_i)$

$$f(x) \cong L(\lambda) = \sum_{i=0}^{N} \lambda_i b_i = \sum_{i=0}^{N} \lambda_i f(a_i) = \sum_{i=0}^{N} \lambda_i f\left(L + \frac{i}{N}(U - L)\right) \tag{30}$$

$$\sum_{i=0}^{N} \lambda_i = 1$$

$$\lambda_i \geq 0 \quad i = 0,1,2, \dots N$$

Phase 6: Define N binary variables with a sum of 1, to determine which of the N intervals x falls into:

$$\sum_{i=1}^{N} y_i = 1 \tag{31}$$

$$y_i \in \{0,1\} \quad i = 1,2, \dots N$$

Phase 7: Add constraints to link $y_i$ to the intervals:

$$\lambda_0 \leq y_1 \tag{32}$$

$$\lambda_1 \leq y_1 + y_2$$

$$\lambda_2 \leq y_2 + y_3$$

$$\lambda_i \leq y_i + y_{i+1} \quad i = 1, \dots N - 1$$

$$\lambda_N \leq y_N$$

This method linearizes the nonlinear function by approximating it with piecewise segments, simplifying optimization and analysis in linear programming. In our model, the nonlinear constraint 3 for travel time on link $l$, $t_l = T_l^0 \left(1 + \alpha \left(\frac{h_l}{V_l}\right)^{\beta}\right)$, includes the term $(h_l)^{\beta}$, which introduces nonlinearity. To linearize this constraint, we apply PLA with a lower bound L $=$ 0 for all $l \in \mathcal{L}$ and an upper bound U $=$ max $(V_l,\ \sum_{o \in \mathcal{O}} \sum_{d \in \mathcal{D}} d^{o \to d})$ $if\ l \in \mathcal{K}_r^{o \to d}$. The range of $h_l$ is divided into segments. First, we discretized $h_l$ by defining breakpoints $a_l^i$, where $a_l^i = L[l] + \frac{i}{N}$(U[l] $-$ L[l]) for $l \in \mathcal{L}$ and i=0, 1, ..., N. For each $a_l^i$, we compute function values $b_l^i$, where $b_l^i =$ $\left(a_l^i\right)^\beta$. The traffic flow $h_l$ is then expressed as a convex combination of these breakpoints: $h_l \approx \sum_{i=0}^{N} \lambda_i a_l^i$, where $\sum_{i=0}^{N} \lambda_i$=1 and $\lambda_i \geq 0$ for i $= 0, 1, \ldots,$ N. To approximate the function range as a convex combination of $b_i = f(a_i)$, we precompute $b_i = (h_l)^\beta$ for each breakpoint, where $(h_l)^\beta \approx \sum_{i=0}^{N} \lambda_i b_i$. Binary variables $y_i$ are introduced to determine the interval where $h_l$ falls, ensuring that exactly one $y_i$ is active, which links $\lambda_i$ and $y_i$ to the corresponding segment. The original nonlinear equation 3 is replaced with:

$$h_l = \sum\nolimits_{i=0}^{N} \lambda_i h_l^i \qquad \forall(l \in \mathcal{L}) \qquad (33)$$

$$\sum\nolimits_{i=0}^{N} \lambda_i = 1 \qquad (34)$$

$$\lambda_i \geq 0 \qquad i = 0, 1, \ldots, N \qquad (35)$$

$$(h_l)^\beta \approx \sum\nolimits_{i=0}^{N} \lambda_i b_i \qquad \forall(l \in \mathcal{L}) \qquad 36)$$

$$\sum_{i=0}^{N} y_i = 1 \qquad y_i = \{0, 1\} \qquad \forall(l \in \mathcal{L}) \qquad (37)$$

$$\lambda_0 \leq y_i, \quad \lambda_i \leq y_i + y_{i+1} \quad , \ldots , \quad \lambda_N \leq y_N \qquad \forall(l \in \mathcal{L}) \qquad (38)$$

Applying PLA to each scenario $h_{l\xi}^{\beta}$in equations 9, 10, and 11 makes the model linear, enabling efficient solutions with linear programming methods while preserving an accurate approximation of the original nonlinear relationship. This approach keeps the model tractable for large-scale optimization while retaining a close approximation of the nonlinear dynamics.

### 4.2. Finding K Shortest Path

The algorithm begins by running Dijkstra's algorithm from the target node to all other nodes in the graph, which allows it to efficiently calculate the shortest path from any node to the target. Using this precomputed shortest path data, the algorithm constructs paths iteratively by combining existing paths with new segments (or "spur paths"). like Yen's algorithm (Yen, 1970), which removes and re-adds edges, Eppstein's algorithm keeps the graph intact and uses a priority queue to manage and explore candidate paths. Each candidate path is evaluated based on its total cost, which is calculated using the edge weights and shortest path distances. The algorithm continues to expand the most promising paths in the priority queue until it finds the top k shortest paths from the source to the target. Pseudocode for Eppstein's Algorithm, which finds K-Shortest K-Shortest loopless paths, is provided in Algorithm 1. This method is particularly efficient, avoiding redundant calculations and enabling rapid retrieval of multiple shortest paths.

Algorithm 1: pseudocode for Eppstein's Algorithm to find K-Shortest loop less paths

```
Input: Graph G, source s, target t, number of paths K
Initialize:
SP ⟵ empty list // List to store the K shortest paths
Q ⟵ empty priority queue // Candidate paths
dist, pred ← Dijkstra(G, t) // Shortest paths from all nodes to t
Q.enqueue((dist[s], s, empty list)) // Enqueue the shortest path cost, start node, and empty path
while Q is not empty and | SP |< K:
    (cost, node, path) ← Q.dequeue() // Dequeue the best candidate path
    if node = t:
        SP.append(backtrackPath(path, t)) // Backtrack to get the full path and add to SP
    else:
        for each neighbor v of node in G:
            if (node, v) is not already in path then:
                newCost ← cost + G[node][v]['weight'] − dist[node] + dist[v]
                newPath ← path + [(node, v)]
                Q.enqueue((newCost, v, newPath)) // Enqueue the new path with updated cost
            end if
        end for
    end if
end while
return SP
```

This approach to scenario-based stochastic programming, enhanced by linearization techniques and efficient pathfinding algorithms, allowed us to solve the problem effectively, yielding valuable insights into the computational complexity and underlying relationships within the model.

## 5. Case study

The case study is based on the Sioux Falls City Road network, which comprises 24 nodes and 76 links, as illustrated in Figure 3. In this model, nodes 1, 2, 3, and 13 are designated as origin points, while nodes 6, 7, 18, and 20 serve as destinations. The corresponding demand values for these origin-destination pairs are detailed in Table 5. Due to the large capacity values within the network, I scaled the original demand by a factor of 10 to better represent traffic flow and make the demand values more manageable for analysis. This adjustment ensures that the traffic model reflects a more realistic scenario, balancing demand with network capacity for improved simulation accuracy. Detailed information about free-flow time and capacity is available at https://github.com/bstabler/TransportationNetworks/tree/master/SiouxFalls.

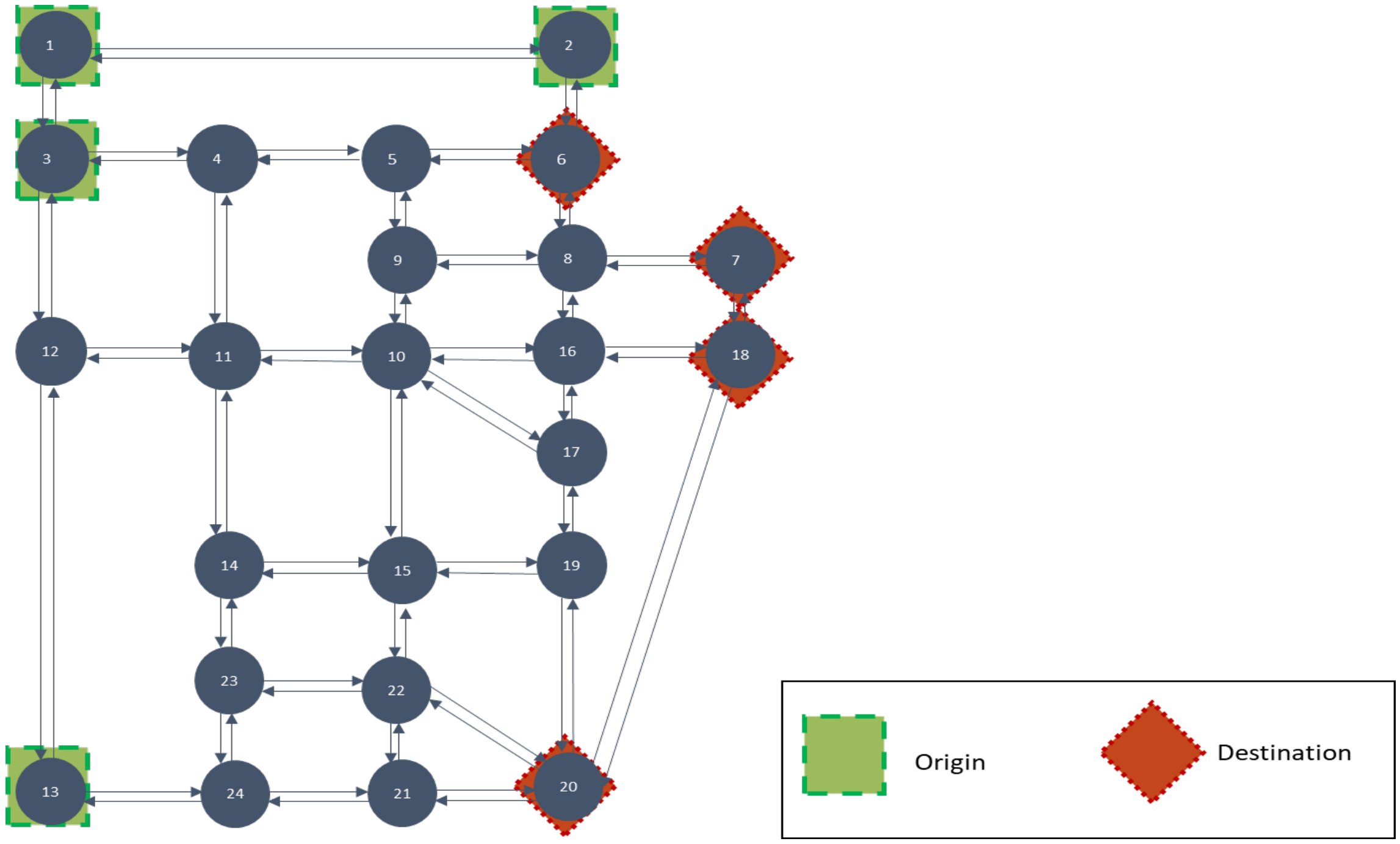

Figure 3. Sioux Falls City Road network

Table 5. Demand

| Origin | Destination | Demand *100 |
|---|---|---|
| 1 | 6 | 100 |
| 2 | 6 | 150 |
| 3 | 6 | 200 |
| 13 | 6 | 250 |
| 1 | 7 | 100 |
| 2 | 7 | 150 |
| 3 | 7 | 200 |
| 13 | 7 | 250 |
| 1 | 18 | 100 |
| 2 | 18 | 150 |
| 3 | 18 | 200 |
| 13 | 18 | 250 |
| 1 | 20 | 100 |
| 2 | 20 | 150 |
| 3 | 20 | 200 |
| 13 | 20 | 250 |

## 5.1. Scenario Representation

The model uses a finite set of scenarios ($\Xi$) with associated probabilities ($\pi$), matching the discrete scenario formulation. Scenarios are generated from the importance of nodes results, determined by 27 local measures. Each disruption scenario specifies which nodes in the network are affected and the extent to which their capacities are reduced as a disruption rate. The impact of these disruptions is modeled for specific nodes and their associated incoming/outgoing links, providing a detailed understanding of potential network vulnerabilities. Sets and indexes, parameters, and decision variables under uncertainty are defined in Table 6 based on some assumptions.

Table 6. Scenario Generation

| **Local Measures** | **Scenario** | **Node: Distribution rate** | **Probability ($\pi$)** |
|---|---|---|---|
| Neighborhood Connectivity | $\xi_1$ | 9: 0.7, 17: 0.65, 16: 0.65, 15: 0.65, 14: 0.65, 19: 0.65, 21: 0.65, 10: 0.65 | **0.08** |
| Phi Node Centrality | $\xi_2$ | 2: 0.7, 6: 0.68, 8: 0.65, 1: 0.64, 7: 0.63, 3: 0.62, 12: 0.61, 18: 0.6 | **0.08** |
| Page Rank | $\xi_3$ | 10: 0.7, 8: 0.68, 11: 0.65, 20: 0.64, 22: 0.63, 16: 0.63, 15: 0.63, 3: 0.62 | **0.08** |
| Harmonic Centrality | $\xi_4$ | 10: 0.7, 11: 0.68, 16: 0.65, 15: 0.64, 8: 0.64, 20: 0.64, 9: 0.63, 22: 0.62 | **0.08** |
| Eigenvector Centrality | $\xi_5$ | 10: 0.7, 15: 0.68, 16: 0.66, 17: 0.64, 22: 0.64, 20: 0.63, 19: 0.62, 11: 0.61 | **0.06** |
| Katz Centrality | $\xi_6$ | 10: 0.7, 15: 0.7, 16: 0.7, 22: 0.7, 11: 0.7, 20: 0.7, 8: 0.7, 17: 0.7 | **0.06** |
| Closeness Centrality | $\xi_7$ | 10: 0.7, 11: 0.69, 16: 0.68, 15: 0.67, 9: 0.66, 17: 0.65, 14: 0.64, 12: 0.63 | **0.06** |
| Betweenness Centrality | $\xi_8$ | 10: 0.7, 11: 0.69, 8: 0.68, 12: 0.67, 16: 0.66, 15: 0.65, 20: 0.64, 4: 0.63 | 0.05 |
| Degree Centrality | $\xi_9$ | 10: 0.7, 11: 0.69, 8: 0.68, 16: 0.68, 15: 0.67, 22: 0.67, 20: 0.65, 3: 0.62 | 0.05 |
| Outdegree Centrality | $\xi_{10}$ | 14: 0.7, 8: 0.69, 10: 0.69, 13: 0.69, 15: 0.69, 22: 0.69, 23: 0.69, 3: 0.68 | 0.03 |
| Outdegree Centrality | $\xi_{11}$ | 14: 0.7, 8: 0.69, 10: 0.69, 13: 0.69, 15: 0.69, 22: 0.69, 23: 0.69, 3: 0.68 | 0.03 |
| Exposure | $\xi_{12}$ | 11: 0.7, 13: 0.7, 17: 0.68, 12: 0.67, 2: 0.66, 19: 0.65, 20: 0.64, 21: 0.64 | 0.03 |
| Aggregate measure | $\xi_{13}$ | 10: 0.7, 11: 0.66, 8: 0.65, 15: 0.64, 20: 0.53, 16: 0.5, 4: 0.48, 3: 0.47 | 0.03 |
| Proportional Flow | $\xi_{14}$ | 10: 0.7, 11: 0.69, 12: 0.68, 8: 0.67, 20: 0.66, 6: 0.65, 18: 0.64, 3: 0.63 | 0.03 |

Table 6 (Continued)

| **Local Measures** | **Scenario** | **Node: Distribution rate** | **Probability ($\pi$)** |
|---|---|---|---|
| Tsallis entropy-based redundancy measure | $\xi_{15}$ | 10: 0.7, 11: 0.69, 8: 0.68, 4: 0.67, 15: 0.66, 3: 0.65, 6: 0.64, 12: 0.63 | 0.03 |
| Star Tsallis entropy-based redundancy | $\xi_{16}$ | 6: 0.7, 2: 0.68, 8: 0.66, 7: 0.64, 18: 0.62, 20: 0.6 | 0.03 |
| Group Centrality | $\xi_{17}$ | 10: 0.7, 11: 0.6, 8: 0.6, 16: 0.6, 15: 0.6, 22: 0.6, 20: 0.6, 3: 0.5 | 0.03 |
| Average Rating | $\xi_{18}$ | 1: 0.7, 3: 0.68, 2: 0.66, 4: 0.64, 6: 0.62, 5: 0.6, 7: 0.4, 8: 0.35 | 0.02 |
| Average Path Distance | $\xi_{19}$ | 1: 0.7, 3: 0.68, 2: 0.66, 4: 0.64, 6: 0.63, 5: 0.6, 7: 0.4, 8: 0.35 | 0.02 |
| Average Path Distance when node $i$ is disrupted | $\xi_{20}$ | 24: 0.7, 23: 0.7, 22: 0.7, 21: 0.7, 20: 0.7, 19: 0.7, 18: 0.7, 17: 0.7 | 0.02 |
| Wighted Node | $\xi_{21}$ | 10: 0.7, 11: 0.68, 8: 0.66, 4: 0.64, 15: 0.64, 3: 0.62, 6: 0.6, 12: 0.58 | 0.02 |
| Weighted Node Measure after a node disruption | $\xi_{22}$ | 12: 0.7, 20: 0.68, 16: 0.66, 17: 0.64, 18: 0.62, 8: 0.6, 11: 0.58, 24: 0.56 | 0.02 |
| Total Undelivered Demand after node j disrupted | $\xi_{23}$ | 1: 0.7, 2: 0.7, 3: 0.7, 6: 0.7, 4: 0.7, 12: 0.7, 5: 0.7, 11: 0.7 | 0.02 |
| Total Changes in the Path distance | $\xi_{24}$ | 1: 0.7, 2: 0.68, 3: 0.66, 6: 0.64, 4: 0.64, 12: 0.6, 5: 0.4, 11: 0.35 | 0.01 |
| Segmentwise | $\xi_{25}$ | 21: 0.7, 20: 0.68, 22: 0.66, 19: 0.64, 23: 0.62, 24: 0.6, 13: 0.58, 14: 0.56 | 0.01 |
| Total Changes in the Path distance | $\xi_{24}$ | 1: 0.7, 2: 0.68, 3: 0.66, 6: 0.64, 4: 0.64, 12: 0.6, 5: 0.4, 11: 0.35 | 0.01 |
| Segmentwise | $\xi_{25}$ | 21: 0.7, 20: 0.68, 22: 0.66, 19: 0.64, 23: 0.62, 24: 0.6, 13: 0.58, 14: 0.56 | 0.01 |
| Complexity Measure-Tsallis | $\xi_{26}$ | 8: 0.7, 4: 0.69, 15: 0.68, 3: 0.67, 6: 0.66, 12: 0.65, 9: 0.64, 20: 0.63 | 0.01 |
| Complexity Measure Distribution | $\xi_{27}$ | 11: 0.7, 8: 0.69, 4: 0.68, 15: 0.67, 3: 0.66, 6: 0.65, 12: 0.64, 9: 0.63 | 0.01 |

In our model, each scenario represents a subset of nodes affected by disruptions and their associated percentage of impact on network links. The set of disruption scenarios includes nodes that vary across different metrics such as Neighborhood Connectivity, Phi Node Centrality, PageRank, and Closeness Centrality, among others. For instance, in scenario ξ1, nodes such as 9, 17, and 16 experience a disruption of 70%, 65%, and 65%, while in scenario ξ2, nodes 2, 6, and 8 face 70%, 68%, and 65% disruption (Figure 4).

The probabilities associated with these scenarios are divided into high-risk (π_high risk) and low-risk (π_low risk) categories. High-risk scenarios, such as $\xi_1$ through $\xi_7$, have higher probabilities (0.08 and 0.06), indicating a greater likelihood of occurrence, while low-risk scenarios ($\xi_8$ through $\xi_{27}$) have smaller probabilities (0.05 to 0.01), reflecting reduced impact. After normalizing the probabilities, the tradeoff between risk categories results in a balanced distribution, ensuring that both high-impact and lower-probability events are considered. This comprehensive analysis of disruption scenarios, node impacts, and probability tradeoffs provides a structured approach to assess network resilience and vulnerability under different disruption conditions.

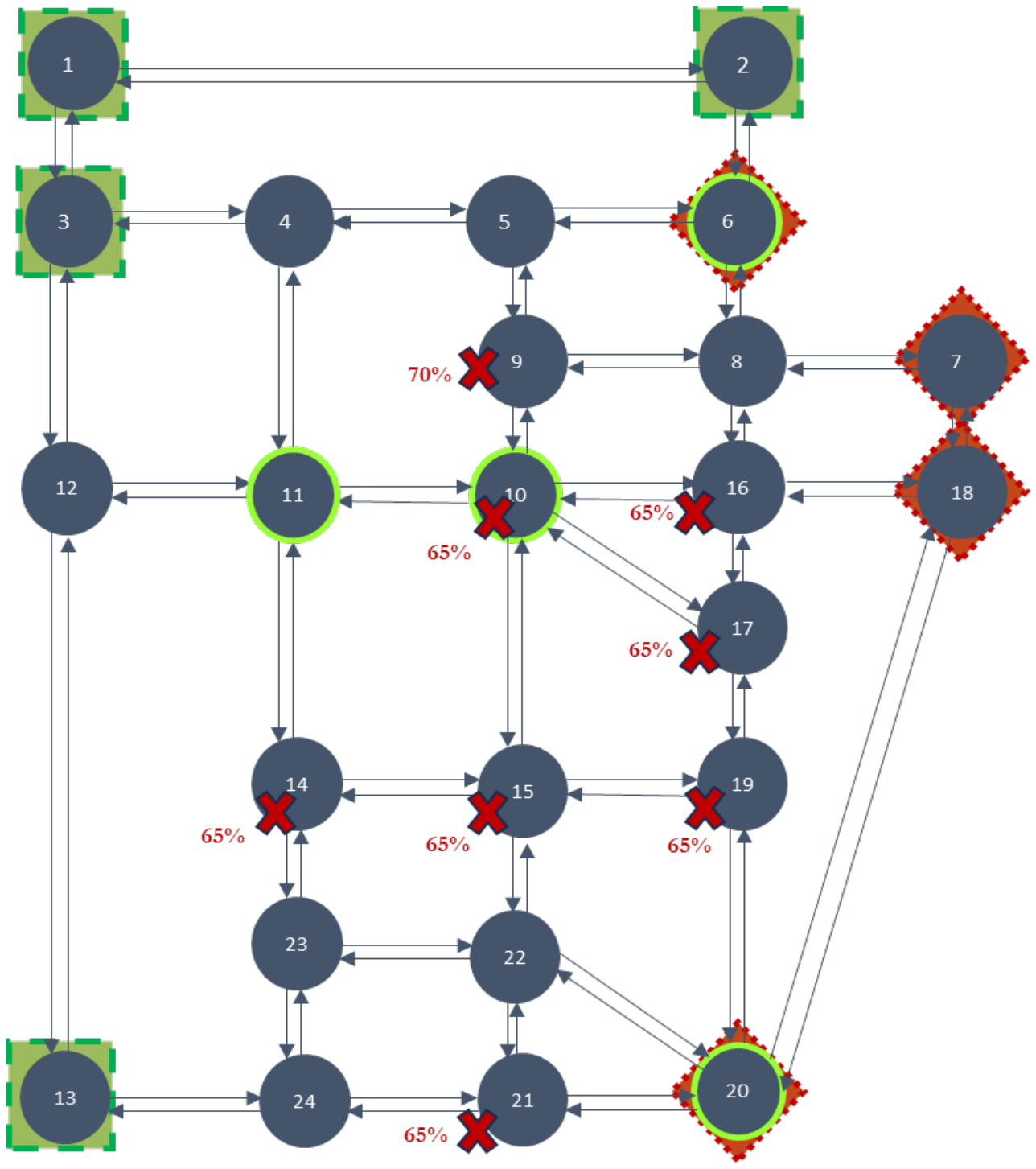

Figure 4. Sioux Falls City Road network under disruption of scenario1

## 6. Results

The analysis of the Sioux Falls City Road Network under both without distribution (normal condition) and with various scenario disruptions (disrupted condition) provides insights into how different models respond. The models considered include the without disruption, expected value (EV), risk neutral two-stage stochastic programming (RN-TSSP), conditional value-at-risk or risk-averse two-stage stochastic programming (CVaR/RA-TSSP), and risk-neutral and risk-averse two-stage stochastic programming (RNRA-TSSP). The key aspects of these models are summarized below:

### 6.1. Optimal Decision Variables in Nominal Conditions (Without Disruption)

Under nominal conditions, the primary decision variable is the optimal flow for each Origin-Destination (O-D) pair, representing the amount of traffic routed through the network to meet demand. The flow values for various O-D pairs and routes, as shown in Figure 5, indicate a flow of 5,000 units for the (13, 20, 9) as (origin, destination, route). Although route 9 is utilized, it does not experience significant congestion compared to the network's busiest routes, such as (13, 7, 6) and (13, 6, 2).

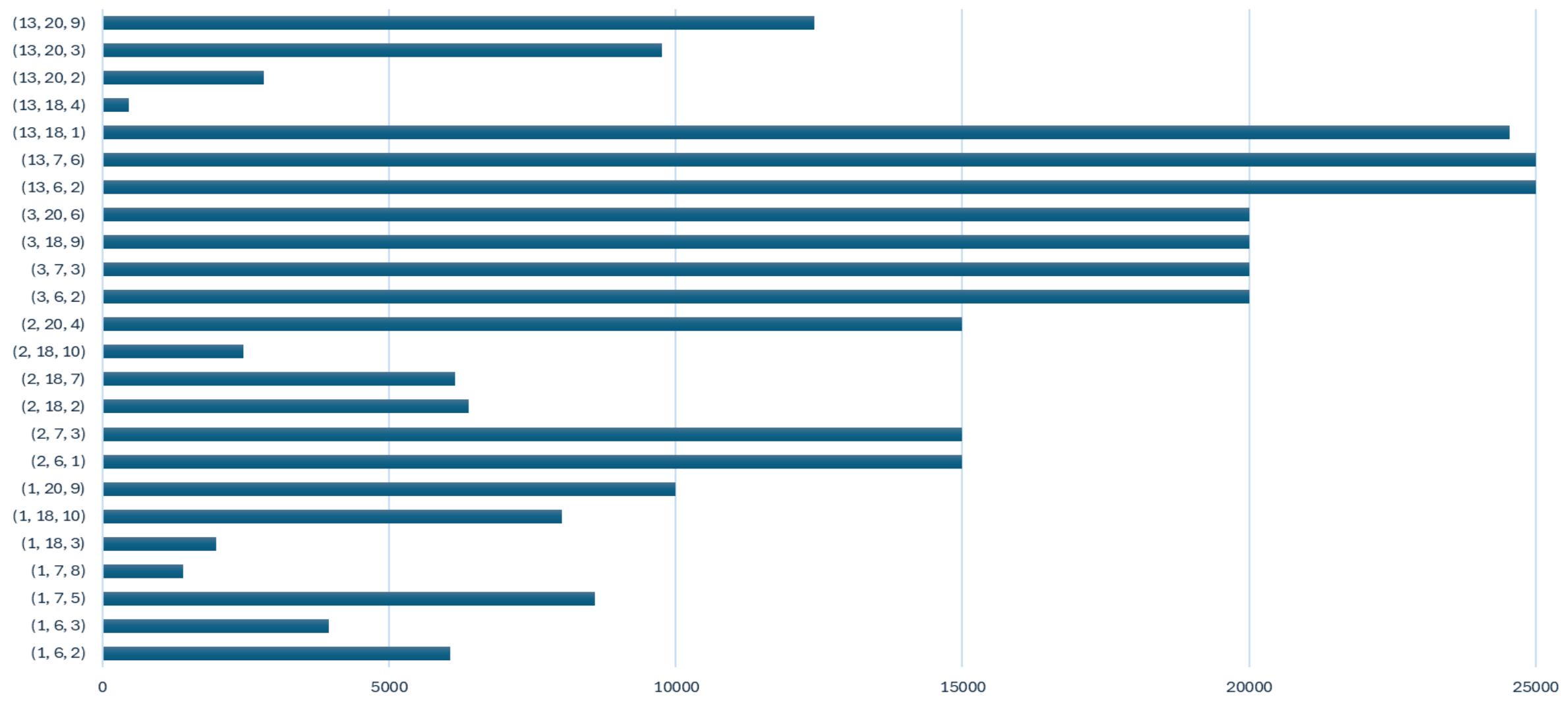

Figure 5. Flow of Sioux Falls City Road network under normal condition (without disruption)

- **O-D Maximum Travel Time**: The Maximum travel time (z) for each O-D pair is determined based on the optimal flow and network capacities, as shown in Table 7. Among all O-D pairs, (3,18) has the lowest Maximum Travel Time due to adequate link capacities, and lower congestion levels along these specific paths makes the travel time.

Table 7. O-D Maximum Travel Time for each O-D pair

| (O, D) | Maximum Travel Time | (O, D) | Maximum Travel Time |
|---|---|---|---|
| (1, 6) | 45082.42 | (2, 18) | 46327.15 |
| (1, 7) | 46323.15 | (2, 20) | 46327.15 |
| (1, 18) | 46327.08 | (3, 6) | 45029.39 |
| (1, 20) | 46320.49 | (3, 7) | 46327.15 |
| (2, 6) | 45089.08 | (3, 18) | **33708.99** |
| (2, 7) | 46327.15 | (3, 20) | 46267.47 |

- **Shortest path:** The K-shortest paths for the Sioux Falls network, presenting multiple viable routes between a given origin and destination, are listed in Appendix J.
- **Undelivered Demand**: In normal conditions, the relative undelivered demand for each O-D pair is minimal or zero, indicating that the network is efficiently handling the demand without significant losses.

### 6.2. RN, RA, and RNRA Model Outputs

In disrupted conditions, the results are compared across three different models: Risk Neutral (RN), Risk Averse (RA) or Conditional Value-at-Risk (CVaR), and a hybrid mean-CVaR model as Risk Neutral Risk Averse (RNRA). Each of these models emphasizes different strategic priorities in terms of fortification, flow distribution, and minimizing undelivered demand. The main metrics of comparison include fortification strategies, network flow, undelivered demand, and maximum travel time for each Origin-Destination (O-D) pair.

#### 6.2.1 Fortification Strategies

As illustrated in Figure 6, The RN model selected nodes 6, 8, 11, and 20 for fortification. The RA model takes a more conservative approach, focusing on minimizing the risk of extreme disruptions. This leads to a more robust fortification plan for worst-case scenarios. In this model, nodes 5, 6, 11, and 20 are selected for fortification, as seen in Figure 7. RNRA (Risk Neutral Risk Averse Hybrid Model): This hybrid model combines the strengths of the RN and RA models, offering a balance between optimizing average performance and mitigating extreme risks. The fortification strategy here, as shown in Figure 8, suggests reinforcing nodes 6, 10, 11, and 20 to effectively manage both average and worst-case conditions.

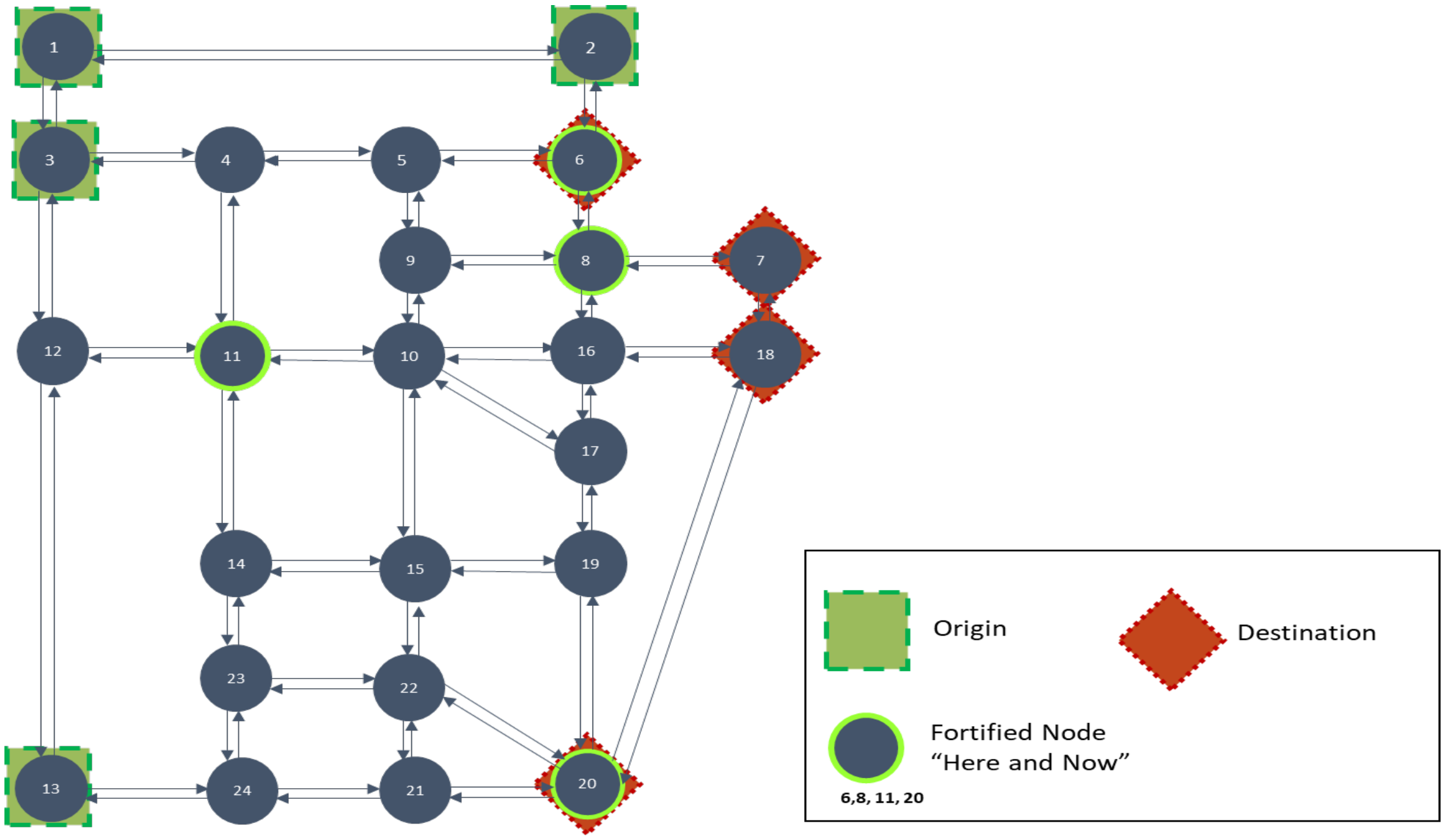

Figure 6. Sioux Falls City Road network based on risk neutral (RN)

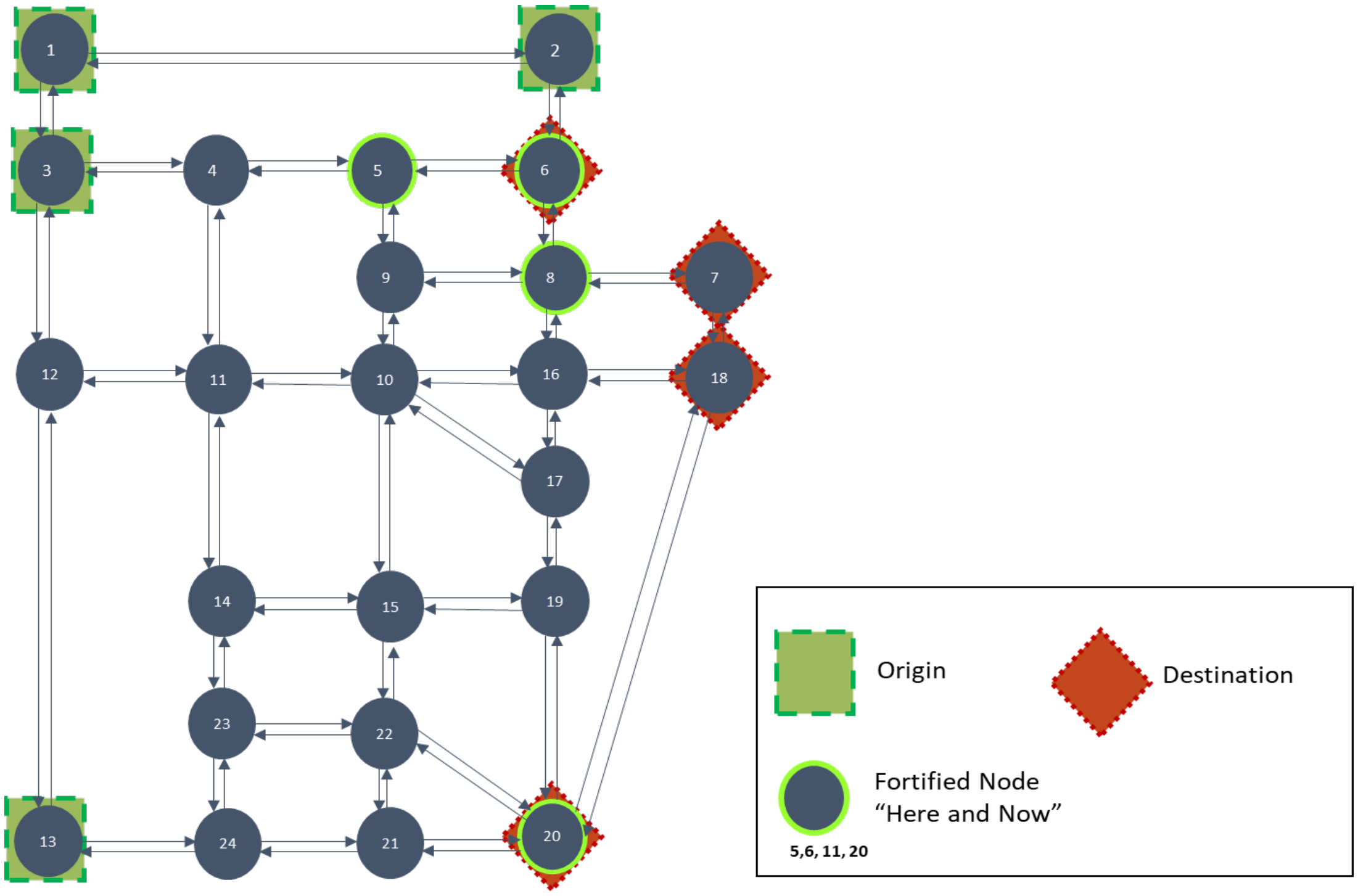

Figure 7. Sioux Falls City Road network based on risk averse (RA)

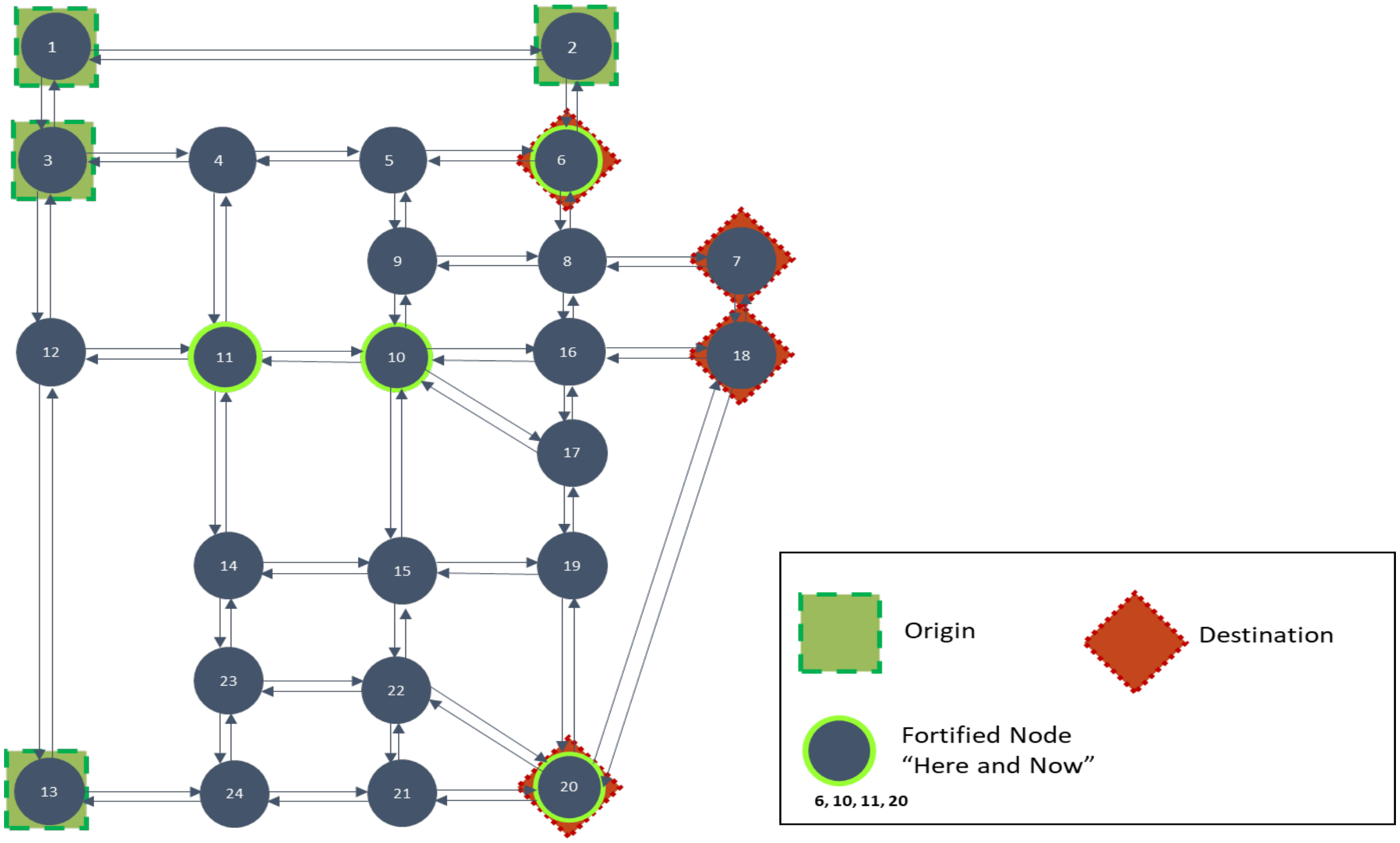

Figure 8. Sioux Falls City Road network based on risk neutral risk averse (RNRA)

### 6.2.2 Undelivered Demand and Travel Time

The RA model generally results in lower undelivered demand compared to the RN model, as it is better equipped to handle worst-case disruptions. The RNRA model strikes a balance between the two, leading to moderate levels of undelivered demand. Figure 9 compares the relatively undelivered demand across different scenarios for the Expected Value, CVaR, and Hybrid approaches, considering the risk level of 0.10. The expected value approach aims to minimize undelivered demand, making it theoretically ideal, but it can show unexpected variations, such as $\xi_{19}$ and $\xi_{23}$. In contrast, CVaR shows greater stability across scenarios but allows for a higher level of unsatisfied demand. The Hybrid strategy offers a balanced approach by minimizing risk while maintaining operational efficiency, making it a more resilient solution overall. In Figure 9 and 10, the results for Delta = 0.30 align closely with expectations, ideal for low-risk operations. Delta = 0.60 offers a balanced approach, mitigating extremes while staying close to expected outcomes, making it suitable for situations where both risk mitigation and operational efficiency are important. Figure 10 highlights that scenarios $\xi_{21}$to $\xi_{24}$ effectively and maximum travel time very efficiently. Finally, Delta = 0.90 prioritizes worst-case scenarios, revealing notable deviations in undelivered demand and travel time, making it ideal for highly risk-averse strategies that focus on worst-case outcomes over expected ones.

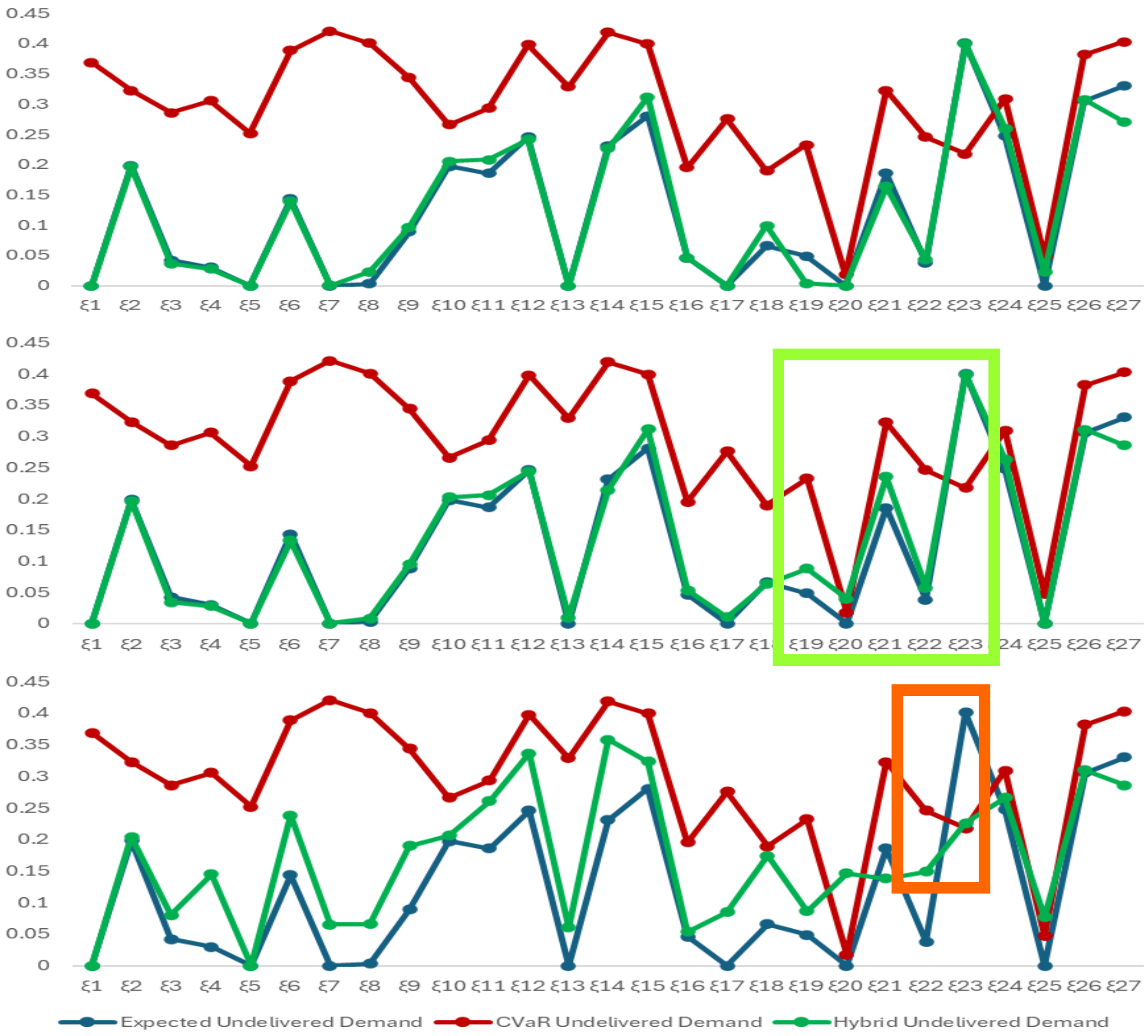

Figure 9. Relative undelivered demand for each scenario based on risk neutral (RN), risk averse (RA), risk neutral risk averse (RNRA) for delta=0.30, 0.60, and 0.90

Similarly, Figure 10 presents the total travel time across scenarios, comparing relative undelivered demand for the Expected value, CVaR, and Hybrid approaches at a risk level of 0.10. The first graph (Delta = 0.30) shows travel time close to expected values, appropriate for low-risk scenarios. The second graph (Delta = 0.60) displays balanced performance between Expected value and CVaR strategies, particularly around $\xi_{17}$to $\xi_{20}$, where variations are moderate, suggesting an optimal trade-off between stability and efficiency. The third graph (Delta = 0.90) aligns more closely with worst-case outcomes, as CVaR dominates with significant deviations in travel time, highlighting that this approach is suited for highly risk-averse strategies that prioritize minimizing extreme risks at the expense of operational efficiency.

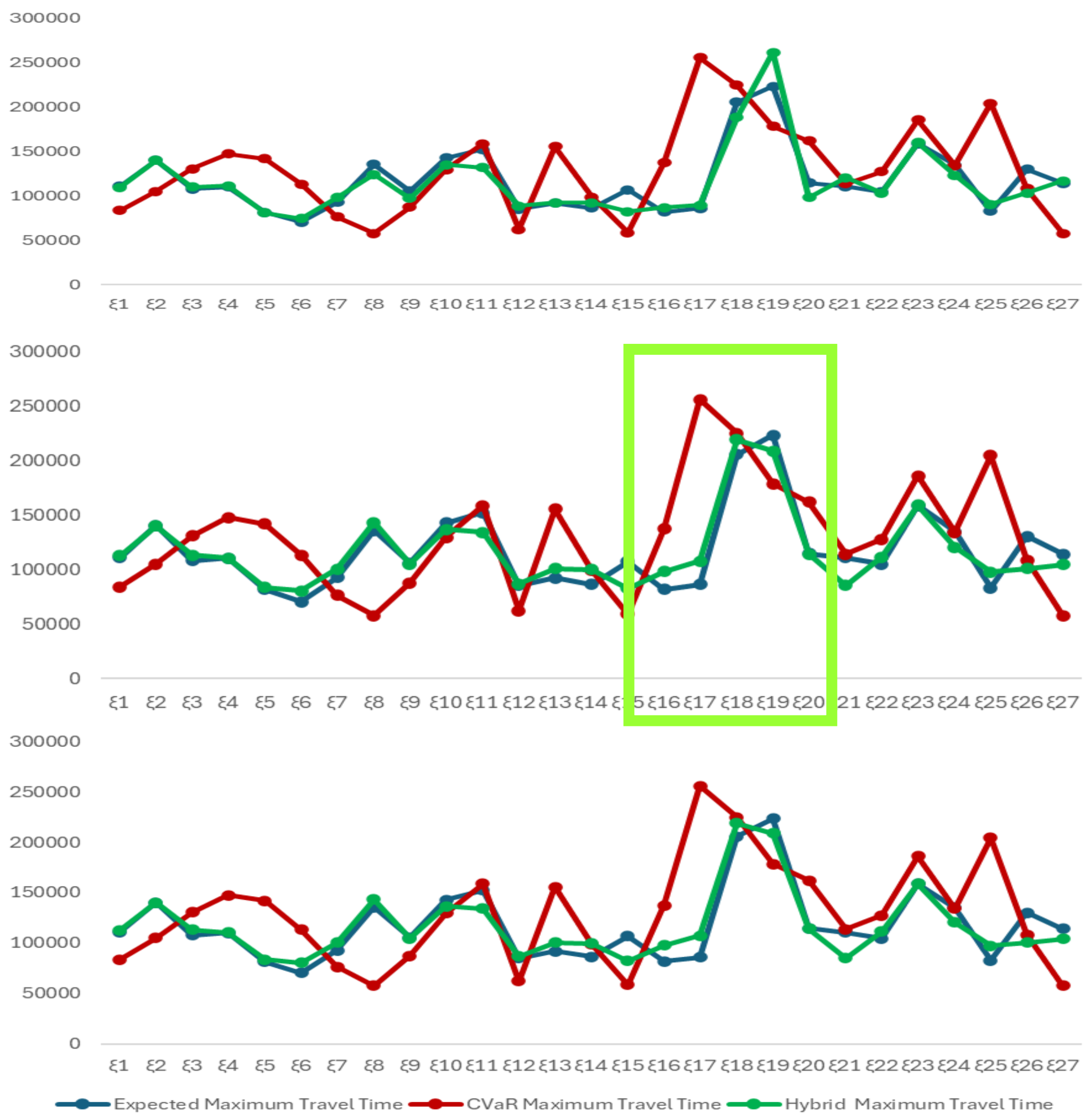

Figure 10. Total travel time for each scenario based on risk neutral (RN), risk averse (RA), risk neutral risk averse (RNRA) for delta=0.30, 0.60, and 0.90

## 7. Discussion and Managerial Insight

Pre-disruption fortifications play a crucial role in reducing risk and enabling the system to maintain normal operations during crises. Each model offers unique strengths—RN targets average disruptions, RA focuses on extreme events, and RNRA approach balances both, providing a comprehensive strategy for managing network performance under different disruption scenarios. Various factors influence planning and risk mitigation, particularly in minimizing undelivered demand, such as the weight assigned to the objective function, higher probability scenarios ($\pi$), and the role of affected nodes. Nodes with multiple connections (both incoming and outgoing links) have a greater impact if disrupted. By applying different levels of fortification, the system demonstrates increased resilience, significantly lowering the risk of failure.

The multi-faceted approach enables more robust planning and risk mitigation strategies, ultimately aiming to minimize undelivered demand across various potential outcomes. In this network scenario, nodes 5, 6, 11, and 20 are fortified in the CVaR model. The average undelivered demand across all scenarios for the expected value, CVaR, and balanced strategies is 0.13, 0.48, and 0.18, respectively. The RNRA model strikes a balance between the extremes of the expected value and CVaR models, demonstrating that the hybrid strategy effectively balances risk, sometimes closely aligning with the expected value model but never reaching the higher levels of undelivered demand seen in the CVaR.

## 8. Conclusion

The compounded effects of natural hazards and infrastructural mismatches underscore the urgent need to rethink current infrastructure development and risk mitigation strategies. Through a comprehensive scenario-based approach, this study highlights how local topological characteristics and centrality measures significantly impact network resilience and vulnerability. The proposed model, which integrates both risk-neutral and risk-averse strategies, provides a robust framework for minimizing objective function outcomes under uncertainty.

The results demonstrate that key factors—such as the probability of disruption, the number of affected nodes, node connectivity, and the resilience of origin-destination paths—play a crucial role in determining the severity of network disruptions and the corresponding undelivered demand. Scenarios with higher probabilities generally result in lower undelivered demand due to improved preparedness, while disruptions to highly connected nodes can trigger cascading failures throughout the network. This highlights the importance of fortifying critical nodes, as their failure can disproportionately affect overall network performance. Our findings emphasize the necessity of a multi-faceted approach to network fortification, which incorporates centrality metrics, risk-aversion strategies (such as CVaR), and a balanced consideration of both high-probability, frequent events and low-probability, high-impact disruptions. By addressing both types of events, decision-makers can develop more comprehensive risk mitigation strategies.

Future research could build upon these insights by refining the model to include more granular spatial and temporal factors, enhancing resilience strategies for complex infrastructure networks in the face of evolving climate conditions. Additionally, employing multi-stage stochastic programming and incorporating socio-economic and demographic variables into the model could further improve demand forecasting and capacity planning. This would enable more precise, tailored, and effective traffic management strategies, ultimately contributing to the creation of more resilient and adaptive infrastructure systems.

## APPENDIX A Local vulnerability measurements

| No. | Measure | Formula | Description |
|---|---|---|---|
| 1 | Complexity measure- Tsallis ($CM_i$) | $CM_i = \frac{A_i}{B_i}$<br>$q_i = (1 + \max BC_i - BC_i)$<br>$A = \left(\frac{\deg i}{\sum_{i \in N} \deg i}\right)^{q_i} - \left(\frac{\deg i}{\sum_{i \in N} \deg i}\right)$<br>$B = 1 - q_i$<br>$q_i$: Difference between the maximum and individual betweenness centrality values | **Complexity measure ($CM_i$):** The complexity degree of the community is measured by Tsallis structure entropy which combines degree distribution and betweenness distribution for each node $i$. |
| 2 | Complexity measure distribution ($CMT_i$) | $CMT_i = C_i$<br>$C = \left(\frac{\deg i}{\sum_{i \in N} \deg i}\right) \log \left(\frac{\deg i}{\sum_{i \in N} \deg i}\right)$ | **Complexity measure distribution:** The distribution of the complexity measure for each node $i$. |
| 3 | Average path distance ($AP_i$) | $AP_i = \frac{2 \sum_{i,j \in OD} d_{ij}}{\lvert OD \rvert}$<br>$\lvert OD \rvert$ the number of origin destination paths<br>$d_{ij}$ = Distance between nodes $i$ and $j$; $i, j \in N$ | **Average path distance:** The node's average connectivity to all other nodes, providing insight into the node's accessibility and its role in facilitating network flow. |
| 4 | Average path distance when node $i$ is disrupted ($APD^{OD}_{\nexists i \in N}$) | The Average Path Distance when node j is disrupted<br>$APD^{OD}_{\nexists i \in N} = \frac{2 \sum_{i,k \in OD, \nexists j \in N} d_{ik}}{\lvert OD \rvert}$<br>$\lvert OD \rvert$ the number of origin destination paths<br>$d_{ij}$ = Distance between nodes $i$ and $j$; $i, j \in N$ | **Average path distance when node $i$ is disrupted:** The sum of distances $d_{ik}$ between all pairs of origin and destination nodes i and k, excluding paths that pass through the disrupted node j, multiplied by 2, indicating how such disruptions might affect the efficiency of flow or transport across the network. |
| 5 | Proportional flow (PF$\varphi_i$) | $\text{PF}\varphi_i = \frac{\text{CF}_i + BC_i}{2}$<br>$BC_i$: Betweenness Centrality | **Proportional flow**: The importance of a node in a network based on the flow passing through it. |
| 6 | Weighted node ($W_i$) | $W_i = \frac{\varphi_i + BC_i}{2}$ | **Weighted node:** The average of $\varphi_i$ and $BC_i$. |
| 7 | Weighted node after a node $j \in N$ disruption ($GW_i$) | $GW_i = \frac{\varphi_{ij} + BC_{ij}}{2} \quad \forall i \in N - \{j\}$ | **Weighted node after a node disruption:** The average |

| No. | Measure | Formula | Description |
|---|---|---|---|
| | | | of $\varphi_i$ and $BC_i$ to find how the disruption of node $j$ impacts the importance of other nodes $i$ in the network. |
| 8 | Total undelivered demand after node j disrupted ($\mathrm{UD}\varphi_\mathrm{i}$) | $\mathrm{UD}\varphi_\mathrm{i} = \sum_{k \in N^+} \eta_{kt_{\nexists j \in N}} - \sum_{k \in N^+} \eta_{kt_0}$<br>$\eta_{jt_0}$ = Total delivered flow to demand node $j$ before disruption $t_0$ | **Total undelivered demand after a node disrupted:** The difference between the total delivered flow after and before the disruption. |
| 9 | Total changes in the path distance ($\Delta \mathrm{d}_\mathrm{i}^\mathrm{OD}$) | $\Delta d_j^{OD} = \bar{d}_j^{OD} - \bar{d}^{OD}$ | **Total change in the path distance:** How much the average path distance between origin-destination pairs changes after a node disruption. |
| 10 | Segmentwise ($seg_i$) | $seg_i = \varphi'_j \Delta d_j^{OD}$ | **Segmentwise**: The impact of node disruption on specific network segments. |

APPENDEX A (Continued)

| No. | Measure | Formula | Description |
|---|---|---|---|
| 11 | Phi node centrality ($\varphi_i$) | $\varphi_i = \frac{f_i}{\sum_{j \in N^+} \eta_{jt_0}}, i \in N\ j \notin N$<br>$\eta_{jt_0}$ = Total delivered flow to demand node $j$ before disruption $t_0$<br>$f_i$ =Total flow passing node $i \in N$ | **Phi node centrality:** The significance of a node i<br>The flow passing through it is normalized by the total delivered flow to all demand nodes before a disruption. It assesses the node's criticality in the network's functionality, highlighting its role in sustaining flow continuity and service delivery. |
| 12 | Degree centrality ($DC_i$) | $DC_i = \frac{\deg_i}{\sum_{j \in N^+} \deg j}$<br>$\deg_i$ = Degree of node $i \in N$ | **Degree centrality:** The number of direct connections that node has to other nodes in the network, reflecting its immediate influence or activity level within the network's structure. |
| 13 | Indegree centrality ($DC_i^+$) | $DC_i^+ = \frac{\deg_i^+}{\sum_{j \in N^+} \deg j}$ | **Indegree centrality:** The number of incoming edges to a node. It is commonly used in directed networks. |
| 14 | Outdegree centrality ($DC_i^-$) | $DC_i^- = \frac{\deg_i^-}{\sum_{j \in N^+} \deg j}$ | **Outdegree centrality:** The number of outgoing edges from a node. It is also used in directed networks. |
| 15 | Betweenness centrality / Load centrality ($BC_i$) | $BC_i = \frac{\sum_{od \in OD} \mu_i^{od}}{\sum_{od \in OD} \mu^{od}}$<br>$\mu_i^{od} = \begin{cases} 1 & if\ the\ shortest\ path\ od\ passes\ node\ i \\ 0 & O.W. \end{cases}$<br>$\mu^{od} = The\ total\ number\ of\ od\ paths$ | **Betweenness centrality / Load centrality:** The number of shortest paths that pass-through a given node. Nodes with high betweenness centrality often serve as bridges or bottlenecks in the network. |
| 16 | Eigenvector centrality ($EI_i$) | $EI_i = \frac{1}{\lambda} \sum_{j \in M_i} x_j = \frac{1}{\lambda} \sum_{j \in N} a_{ij} x_j$<br>$a_{ij}$ = The adjacency matrix element | **Eigenvector centrality:** The importance of a node based on the importance of its neighbors. A node with high eigenvector centrality is connected to other nodes with high centrality. |
| 17 | Katz centrality ($KC_i$) | $KC_i = \sum_{k=1}^{\infty} \sum_{j=1}^{N} \alpha^k (A^k)_{ij}$<br>$\alpha = 0.5$<br>$A$ = the adjacency matrix | **Katz centrality:** The influence of a node in a network is based on the number of paths that pass through it. It is similar to eigenvector |
| | | | centrality but considers paths of different lengths. |
| 18 | Closeness centrality ($CC_i$) | $CC_i = \frac{1}{\sum_{j \in N} d_{ij}}$<br>$d_{ij}$ = Distance between nodes $i$ and $j$ $i, j \in N$<br>The Harmonic Centrality | **Closeness centrality:** The average length of the shortest paths from a node to all other nodes in the network. Nodes with high closeness centrality are often highly connected. |
| 19 | Harmonic centrality ($HC_i$) | $HC_i = \sum_{j \in N} \frac{1}{d_{ij}}$<br>$d_{ij}$ = Distance between nodes $i$ and $j$ $i, j \in N$ | **Harmonic centrality:** The average distance of a node to all other nodes in the network. It is a variant of closeness centrality that sums the inverse of the geodesic distances of each node to other nodes where it is connected. |
| 20 | Aggregate measure ($AG_i$) | $AG_i = \frac{w_1(DC_i - \overline{DC})}{\overline{DC}} + \frac{w_2(He_i - \overline{He})}{\overline{He}} + \frac{w_3(BC_i - \overline{BC})}{\overline{BC}}$<br>$w_1$, $w_2$ , and $w_3$ are weights for each centrality measure. | **Aggregate measure:** The overall centrality by standardizing the centrality measures. |
| 21 | Neighborhood connectivity ($NC_i$) | $NC_i = \frac{\sum_{i \in N} \frac{\sum_{\exists(j,i) \vee \exists(i,j) \in L} \deg j}{\sum_{j \in N} a_{ij}}}{\lvert N \rvert}$ | **Neighborhood connectivity:** The average connectivity of all nodes directly connected to node $i$. |

(Continued)

| No. | Measure | Formula | Description |
|---|---|---|---|
| 22 | Star Tsallis entropy-based redundancy measure ($TE_i$) | $TE_i = \frac{1}{e_m - 1} \sum_{j \in N} Q_{ij}$<br>$e_m = 1.43$<br>$Q_{ij} = \left(\frac{f_{ij}}{f_i}\right) - \left(\frac{f_{ij}}{f_i}\right)^2$ | **Tsallis entropy-based redundancy measure, or shannon entropy**: The redundancy present in the layouts of water distribution networks, which refers to the existence of multiple paths or connections that can ensure the flow of water even in the case of failures or disruptions |
| 23 | Tsallis entropy-based redundancy measure ($TE_i^*$) | $TE_i^* = \frac{1}{e_m - 1}\left(\left(\frac{\sum_{j \in N} f_{ij}}{\sum_{i \in N} f_i}\right) - Q_i^*\right)$<br>$e_m = 1.43$<br>$Q_i^* = \sum_{j \in N}\left(\frac{f_{ij}}{\sum_{i \in N} f_i}\right)$ | |
| 24 | Page rank ($PR_i$) | $PR_i = \frac{PR_j}{\sum_{j \in N} a_{ij}}$<br>$PR_j$: is the PageRank of page j<br>$\sum_{j \in N} a_{ij}$ : $N$ormalized by their number of outbound edges | **Page rank**: An algorithm used by web search engines to rank web pages in their search results, where $PR_i$ represents the PageRank of page i, quantifying its importance based on the quantity and quality of links pointing to it from other pages. |
| 25 | Exposure ($Expo_i$) | $Expo_i = \frac{d_{kl,\nexists i \in N} - d_{kl}}{\lvert i \rvert \lvert N \rvert (\lvert N \rvert - 1)}$<br>$\lvert i \rvert$ The exposure of the $i^{th}$ disrupted node | **Exposure:** The impact of removing node $i$ on the distances between other nodes, providing insights into the node's influence on network connectivity. |
| 26 | Group centrality ($GrC_i$) | $GrC_i = \sum_{j \in N, \nexists (i,j) \in L \vee (j,i) \in L} BC_j$ | **Group centrality:** The influence or importance of nodes that are indirectly connected to node $i$ through other paths in the network. Nodes with higher Group Centrality Measure values are considered more influential or central in the network. |
| 27 | Average rating ($AR_i$) | $AR_i = \frac{1}{\lvert N \rvert}\left(\frac{1}{\min_{i \in N} \deg i}\right)$ | **Average rating:** Assess the average rating or importance of nodes in a network based on their minimum degree. Nodes with higher |
| | | | minimum degrees are considered more important or influential in the network. |

APPENDIX B Summary of vulnerability local measurements in transportation, power, water, and community

| No. | Study (year) | LM Label | Application | Topology | Attack |
|---|---|---|---|---|---|
| 1 | N. Wang et al. (2023) | $DC_i, BC_i, CC_i, EI_i$ | Bus, Subway, Taxi | Complex | Random, Target |
| 2 | Yin et al. (2023) | $DC_i, BC_i, CC_i$ | Railway | Space-L | Random, Target |
| 3 | Ferrari & Santagata (2023) | $DC_i, BC_i, CC_i, EI_i$ | Motorways, Railways | 13 pure topologies | Random, Target |
| 4 | Huang et al. (2021) | $\varphi_i$ | Metro | Scale-free | Random |
| 5 | Klophaus et al. (2021) | $DC_i, BC_i$ | Airline | Mesh | - |
| 6 | Zhou et al. (2021) | $BC_i$ | Airport | Random, Scale-free, | - |
| 7 | Almotahari & Yazici, (2021) | LClij , MCij ,Rdij | Road | Nguyen-Dupuis, Sioux Falls, Anaheim | - |
| 8 | Zhao & Xiu (2021) | $BC_i, CC_i$ | US Air Cargo | Random, Small-world, Scale-free | Target |
| 9 | Liu et al., (2020) | $BC_i, HC_i, DC_i, AG_i$ | Rail | Complex | - |
| 10 | Lafia-Bruce (2020) | $BC_i, CC_i, EI_i$ | Road | Complex | - |
| 11 | Cats & Krishnakumari (2020) | $BC_i, DC_i$ | Rail | Complex | Random, Target |
| 12 | Peng et al. (2019) | $BC_i, DC_i, CC_i$ | Crude oil transport (Interdependency with) | Hub & spoke | - |
| 13 | Akbarzadeh et al. (2019) | $BC_i, DC_i$ | Road | Complex | Target |

(Continued)

| No. | Study (year) | LM Label | Application | Topology | Attack |
|---|---|---|---|---|---|
| 14 | Xiao et al. (2019) | $BC_i$ | Metro | Space-L | - |
| 15 | Q.-C. Lu, (2018) | $BC_i$ | Rail | Complex | - |
| 16 | L. Sun et al. (2018) | $DC_i, BC_i$ | Rail | Complex | Target |
| 17 | Zanin et al. (2018) | $BC_i, CC_i, EI_i, PR_i, KC_i$ | Air, Bus, Light rail, Subway, and Tram | Complex | Random, Target |
| 18 | Peng et al. (2018) | $DC_i, BC_i$ | Cargo | Complex | Random, Target |
| 19 | H. Zhang et al. (2018) | $DC_i, BC_i, NC_i$ | Bus | L-space, P-space | Target |
| 20 | Cats (2017a) | $DC_i, CC_i, BC_i$ | Rail | L-space | - |
| 21 | López et al. (2017) | $DC_i, CC_i, BC_i, EI_i$ | Transport | Point-to-point, Hub & spoke, Random | Random |
| 22 | Lordan et al. (2016) | $DC_i, BC_i$ | Flight | Point-to-point, Hub & spoke | Random, Target |
| 23 | D. J. Sun & Guan (2016) | $BC_i$ | Metro | Complex, Scale-free, | Random |
| | | | | Small-world, Random | |
| 24 | Yang et al. (2015) | $DC_i + BC_i$ | Subway | Scale-free | Random, Target |
| 25 | Salama et al. (2022) | $DC_i, BC_i$ | Power grid | Small-world, Mesh | Random, Target |
| 26 | Li et al. (2021) | $DC_i, BC_i$ | Power grid | Complex | Target |
| 27 | Galindo-González et al. (2020) | $DC_i, BC_i$ | Power grid | Complex | Target, Random |
| 28 | Panigrahi & Maity (2020) | $DC_i, BC_i$ | Power grid | Scale-free | Target |
| 29 | Albarakati (2020) | $DC_i, BC_i, EI_i$ | Power grid | Scale-free, Small-world | Target |
| 30 | Reyes et al. (2019) | $HC_i, CC_i$ | Microgrid | Circular, Double-tree | - |

(Continued)

| No. | Study (year) | LM Label | Application | Topology | Attack |
|---|---|---|---|---|---|
| 31 | Shahpari et al. (2019) | $DC_i, BC_i$ | Power grid | Complex | - |
| 32 | Albarakati & Bikdash (2018) | $DC_i$ | Transmission lines | Scale-free, Small-world | Random |
| 33 | Faramondi et al. (2018) | $DC_i, BC_i, EI_i$ | Power grid, Air transportation | - | Target |
| 34 | Espejo et al. (2018) | $BC_i$ | Transmission grid | Small-world, Mesh | - |
| 35 | Cetinay et al., (2018) | $DC_i, BC_i, EI_i, CC_i$ | Power grid | Complex | Target |
| 36 | Guo et al. (2018) | $DC_i, BC_i, CC_i, EC_i$ | Power grid | Small world | Target, Random |
| 37 | Warnier et al. (2017) | $EN_i$ | Power grid | Grid | Target |
| 38 | Kim et al. (2017) | $DC_i, BC_i$ | Power grid | Complex: Korean power Grid, Random, Scale-free | Target, Random |
| 39 | Xue et al. (2017) | $BC_i$ | Power grid | Complex | Target, Random |
| 40 | Mureddu et al. (2016) | $BC_i$ | Transmission grid | Complex | Target, Random |
| 41 | Rout et al. (2016) | $BC_i$ | Power grid | Complex | - |
| 42 | Bhave et al. (2016) | $BC_i, CC_i, DC_i$ | Power grid | Complex | Target |
| 43 | Caro-Ruiz & Mojica-Nava (2015) | $DC_i, CC_i, EI_i$ | Power grid | Complex | - |
| 44 | Santonastaso et al. (2021) | $BC_i$ | Water distribution | Complex | - |
| 45 | Ponti et al. (2021) | $BC_i, CC_i, DC_i$ | Water distribution | Complex | - |

(Continued)

| No. | Study (year) | LM Label | Application | Topology | Attack |
|---|---|---|---|---|---|
| 46 | Mortula et al. (2020) | $DC_i, DC_i^+, DC_i^-$ $CC_i, BC_i, EI_i, HC_i$ | Water distribution | Complex | - |
| 47 | Wéber et al. (2020) | $APD_{\nexists i \in N}^{OD}, \Delta d_i^{OD}, seg_i, UD\varphi_i$ | Water distribution | Random, Scale-free | - |
| 48 | Giustolisi et al. (2020) | $BC_i, HC_i, DC_i$ | Water distribution | Complex | - |
| 49 | F. Wang et al. (2019) | $W_i, GW_i$ | Water distribution | Complex | - |
| 50 | Zarghami & Gunawan (2019) | $CC_i, BC_i, EI_i$ | Water distribution | Complex | - |
| 51 | Xu et al. (2019) | $DC_i, BC_i$ | Water distribution | Small-world | - |
| 52 | Nazempour et al. (2018) | $CC_i, DC_i, BC_i$ | Water distribution | Complex | Target |
| 53 | Mazumder et al. (2020) | $CC_i, DC_i, BC_i$ | Water distribution | Complex | - |
| 54 | Agathokleous et al. (2017) | $BC_i$ | Water distribution | Complex | - |
| 55 | Z. Liu et al. (2015) | $CC_i, DC_i, BC_i$ | Water distribution | Complex | Target |
| 56 | Singh & Oh (2015) | $CM_i, CMT_i$ | Water distribution | Complex | - |
| 57 | Parimi et al. (2022) | $DC_i, BC_i, EI_i, CC_i$ | Power grid, social network, Railway | Complex | Target |
| 58 | Wen & Deng (2020) | $DC_i, BC_i$ | Telephone, Social network, Email, Power transmission grid | Complex | Random |
| 59 | Wang et al. (2019) | $BC_i$ | Subway, Power grid | Small-world, Scale-free, Random | Target, Random |
| | | | | | |

(Continued)

| No. | Study (year) | LM Label | Application | Topology | Attack |
|---|---|---|---|---|---|
| 60 | Wei et al. (2018) | $DC_i$ | Power transmission grid, social network, Air transportation | Complex | Target |
| 61 | Lujak & Giordani (2018) | $DC_i, EI_i, BC_i$ | Evacuation | Complex | - |
| 62 | Vogiatzis & Pardalos (2016) | $GrC_i$ | Evacuation | Complex | - |
| 63 | Ghedini & Ribeiro (2015) | $BC_i$ | Transportation | Complex (Scale-free) | Target |
| 64 | Abdulla & Birgisson (2020) | $DC_i, EI_i, BC_i, CC_i$ | Transportation | - | Target, Random |

APPENDIX C Shortest path based on Eppstein's K shortest path under normal conditions

| No | (origin, destination, route number) | Shortest paths |
|---|---|---|
| 1 | (1, 6, 1) | [(1, 2), (2, 6)] |
| 2 | (1, 6, 2) | [(1, 3), (3, 4), (4, 5), (5, 6)] |
| 3 | (1, 6, 3) | [(1, 3), (3, 4), (4, 5), (5, 9), (9, 8), (8, 6)] |
| 4 | (1, 6, 4) | [(1, 3), (3, 12), (12, 11), (11, 4), (4, 5), (5, 6)] |
| 5 | (1, 6, 5) | [(1, 3), (3, 4), (4, 11), (11, 10), (10, 9), (9, 5), (5, 6)] |
| 6 | (1, 6, 6) | [(1, 3), (3, 4), (4, 11), (11, 10), (10, 9), (9, 8), (8, 6)] |
| 7 | (1, 6, 7) | [(1, 3), (3, 4), (4, 11), (11, 10), (10, 16), (16, 8), (8, 6)] |
| 8 | (1, 6, 8) | [(1, 3), (3, 12), (12, 11), (11, 10), (10, 9), (9, 5), (5, 6)] |
| 9 | (1, 6, 9) | [(1, 3), (3, 12), (12, 11), (11, 10), (10, 9), (9, 8), (8, 6)] |
| 10 | (1, 6, 10) | [(1, 3), (3, 12), (12, 11), (11, 10), (10, 16), (16, 8), (8, 6)] |
| 11 | (1, 7, 1) | [(1, 2), (2, 6), (6, 8), (8, 7)] |
| 12 | (1, 7, 2) | [(1, 2), (2, 6), (6, 5), (5, 9), (9, 8), (8, 7)] |
| 13 | (1, 7, 3) | [(1, 2), (2, 6), (6, 8), (8, 16), (16, 18), (18, 7)] |
| 14 | (1, 7, 4) | [(1, 3), (3, 4), (4, 5), (5, 6), (6, 8), (8, 7)] |
| 15 | (1, 7, 5) | [(1, 3), (3, 4), (4, 5), (5, 9), (9, 8), (8, 7)] |
| 16 | (1, 7, 6) | [(1, 3), (3, 4), (4, 11), (11, 10), (10, 9), (9, 8), (8, 7)] |
| 17 | (1, 7, 7) | [(1, 3), (3, 4), (4, 11), (11, 10), (10, 16), (16, 8), (8, 7)] |
| 18 | (1, 7, 8) | [(1, 3), (3, 4), (4, 11), (11, 10), (10, 16), (16, 18), (18, 7)] |
| 19 | (1, 7, 9) | [(1, 3), (3, 12), (12, 11), (11, 10), (10, 9), (9, 8), (8, 7)] |
| 20 | (1, 7, 10) | [(1, 3), (3, 12), (12, 11), (11, 10), (10, 16), (16, 8), (8, 7)] |
| 21 | (1, 18, 1) | [(1, 2), (2, 6), (6, 8), (8, 7), (7, 18)] |
| 22 | (1, 18, 2) | [(1, 2), (2, 6), (6, 8), (8, 16), (16, 18)] |
| 23 | (1, 18, 3) | [(1, 3), (3, 4), (4, 11), (11, 10), (10, 16), (16, 18)] |
| 24 | (1, 18, 4) | [(1, 3), (3, 12), (12, 11), (11, 10), (10, 16), (16, 18)] |
| 25 | (1, 18, 5) | [(1, 2), (2, 6), (6, 5), (5, 9), (9, 8), (8, 7), (7, 18)] |
| 26 | (1, 18, 6) | [(1, 2), (2, 6), (6, 5), (5, 9), (9, 8), (8, 16), (16, 18)] |
| 27 | (1, 18, 7) | [(1, 2), (2, 6), (6, 5), (5, 9), (9, 10), (10, 16), (16, 18)] |
| 28 | (1, 18, 8) | [(1, 2), (2, 6), (6, 8), (8, 9), (9, 10), (10, 16), (16, 18)] |
| 29 | (1, 18, 9) | [(1, 3), (3, 4), (4, 5), (5, 6), (6, 8), (8, 7), (7, 18)] |
| 30 | (1, 18, 10) | [(1, 3), (3, 4), (4, 5), (5, 6), (6, 8), (8, 16), (16, 18)] |
| 31 | (1, 20, 1) | [(1, 2), (2, 6), (6, 8), (8, 7), (7, 18), (18, 20)] |
| 32 | (1, 20, 2) | [(1, 2), (2, 6), (6, 8), (8, 16), (16, 18), (18, 20)] |
| 33 | (1, 20, 3) | [(1, 3), (3, 12), (12, 13), (13, 24), (24, 21), (21, 20)] |
| 34 | (1, 20, 4) | [(1, 2), (2, 6), (6, 8), (8, 16), (16, 17), (17, 19), (19, 20)] |
| 35 | (1, 20, 5) | [(1, 3), (3, 4), (4, 11), (11, 10), (10, 15), (15, 19), (19, 20)] |
| 36 | (1, 20, 6) | [(1, 3), (3, 4), (4, 11), (11, 10), (10, 15), (15, 22), (22, 20)] |
| 37 | (1, 20, 7) | [(1, 3), (3, 4), (4, 11), (11, 10), (10, 16), (16, 18), (18, 20)] |

(Continued)

| No | (origin, destination, route number) | Shortest paths |
|---|---|---|
| 38 | (1, 20, 8) | [(1, 3), (3, 4), (4, 11), (11, 10), (10, 17), (17, 19), (19, 20)] |
| 39 | (1, 20, 9) | [(1, 3), (3, 4), (4, 11), (11, 14), (14, 15), (15, 19), (19, 20)] |
| 40 | (1, 20, 10) | [(1, 3), (3, 4), (4, 11), (11, 14), (14, 15), (15, 22), (22, 20)] |
| 41 | (2, 6, 1) | [(2, 6)] |
| 42 | (2, 6, 2) | [(2, 1), (1, 3), (3, 4), (4, 5), (5, 6)] |
| 43 | (2, 6, 3) | [(2, 1), (1, 3), (3, 4), (4, 5), (5, 9), (9, 8), (8, 6)] |
| 44 | (2, 6, 4) | [(2, 1), (1, 3), (3, 12), (12, 11), (11, 4), (4, 5), (5, 6)] |
| 45 | (2, 6, 5) | [(2, 1), (1, 3), (3, 4), (4, 11), (11, 10), (10, 9), (9, 5), (5, 6)] |
| 46 | (2, 6, 6) | [(2, 1), (1, 3), (3, 4), (4, 11), (11, 10), (10, 9), (9, 8), (8, 6)] |
| 47 | (2, 6, 7) | [(2, 1), (1, 3), (3, 4), (4, 11), (11, 10), (10, 16), (16, 8), (8, 6)] |
| 48 | (2, 6, 8) | [(2, 1), (1, 3), (3, 12), (12, 11), (11, 10), (10, 9), (9, 5), (5, 6)] |
| 49 | (2, 6, 9) | [(2, 1), (1, 3), (3, 12), (12, 11), (11, 10), (10, 9), (9, 8), (8, 6)] |
| 50 | (2, 6, 10) | [(2, 1), (1, 3), (3, 12), (12, 11), (11, 10), (10, 16), (16, 8), (8, 6)] |
| 51 | (2, 7, 1) | [(2, 6), (6, 8), (8, 7)] |
| 52 | (2, 7, 2) | [(2, 6), (6, 5), (5, 9), (9, 8), (8, 7)] |
| 53 | (2, 7, 3) | [(2, 6), (6, 8), (8, 16), (16, 18), (18, 7)] |
| 54 | (2, 7, 4) | [(2, 1), (1, 3), (3, 4), (4, 5), (5, 6), (6, 8), (8, 7)] |
| 55 | (2, 7, 5) | [(2, 1), (1, 3), (3, 4), (4, 5), (5, 9), (9, 8), (8, 7)] |
| 56 | (2, 7, 6) | [(2, 6), (6, 5), (5, 9), (9, 8), (8, 16), (16, 18), (18, 7)] |
| 57 | (2, 7, 7) | [(2, 6), (6, 5), (5, 9), (9, 10), (10, 16), (16, 8), (8, 7)] |
| 58 | (2, 7, 8) | [(2, 6), (6, 5), (5, 9), (9, 10), (10, 16), (16, 18), (18, 7)] |
| 59 | (2, 7, 9) | [(2, 6), (6, 8), (8, 9), (9, 10), (10, 16), (16, 18), (18, 7)] |
| 60 | (2, 7, 10) | [(2, 1), (1, 3), (3, 4), (4, 11), (11, 10), (10, 9), (9, 8), (8, 7)] |
| 61 | (2, 18, 1) | [(2, 6), (6, 8), (8, 7), (7, 18)] |
| 62 | (2, 18, 2) | [(2, 6), (6, 8), (8, 16), (16, 18)] |
| 63 | (2, 18, 3) | [(2, 6), (6, 5), (5, 9), (9, 8), (8, 7), (7, 18)] |
| 64 | (2, 18, 4) | [(2, 6), (6, 5), (5, 9), (9, 8), (8, 16), (16, 18)] |
| 65 | (2, 18, 5) | [(2, 6), (6, 5), (5, 9), (9, 10), (10, 16), (16, 18)] |
| 66 | (2, 18, 6) | [(2, 6), (6, 8), (8, 9), (9, 10), (10, 16), (16, 18)] |
| 67 | (2, 18, 7) | [(2, 1), (1, 3), (3, 4), (4, 11), (11, 10), (10, 16), (16, 18)] |
| 68 | (2, 18, 8) | [(2, 1), (1, 3), (3, 12), (12, 11), (11, 10), (10, 16), (16, 18)] |
| 69 | (2, 18, 9) | [(2, 6), (6, 5), (5, 4), (4, 11), (11, 10), (10, 16), (16, 18)] |
| 70 | (2, 18, 10) | [(2, 6), (6, 5), (5, 9), (9, 10), (10, 17), (17, 16), (16, 18)] |
| 71 | (2, 20, 1) | [(2, 6), (6, 8), (8, 7), (7, 18), (18, 20)] |
| 72 | (2, 20, 2) | [(2, 6), (6, 8), (8, 16), (16, 18), (18, 20)] |
| 73 | (2, 20, 3) | [(2, 6), (6, 8), (8, 16), (16, 17), (17, 19), (19, 20)] |
| 74 | (2, 20, 4) | [(2, 1), (1, 3), (3, 12), (12, 13), (13, 24), (24, 21), (21, 20)] |
| 75 | (2, 20, 5) | [(2, 6), (6, 5), (5, 9), (9, 8), (8, 7), (7, 18), (18, 20)] |
| 76 | (2, 20, 6) | [(2, 6), (6, 5), (5, 9), (9, 8), (8, 16), (16, 18), (18, 20)] |
| 77 | (2, 20, 7) | [(2, 6), (6, 5), (5, 9), (9, 10), (10, 15), (15, 19), (19, 20)] |

Table J.1 (Continued)

| No | (origin, destination, route number) | Shortest paths |
|---|---|---|
| 78 | (2, 20, 8) | [(2, 6), (6, 5), (5, 9), (9, 10), (10, 15), (15, 22), (22, 20)] |
| 79 | (2, 20, 9) | [(2, 6), (6, 5), (5, 9), (9, 10), (10, 16), (16, 18), (18, 20)] |
| 80 | (2, 20, 10) | [(2, 6), (6, 5), (5, 9), (9, 10), (10, 17), (17, 19), (19, 20)] |
| 81 | (3, 6, 1) | [(3, 1), (1, 2), (2, 6)] |
| 82 | (3, 6, 2) | [(3, 4), (4, 5), (5, 6)] |
| 83 | (3, 6, 3) | [(3, 4), (4, 5), (5, 9), (9, 8), (8, 6)] |
| 84 | (3, 6, 4) | [(3, 12), (12, 11), (11, 4), (4, 5), (5, 6)] |
| 85 | (3, 6, 5) | [(3, 4), (4, 11), (11, 10), (10, 9), (9, 5), (5, 6)] |
| 86 | (3, 6, 6) | [(3, 4), (4, 11), (11, 10), (10, 9), (9, 8), (8, 6)] |
| 87 | (3, 6, 7) | [(3, 4), (4, 11), (11, 10), (10, 16), (16, 8), (8, 6)] |
| 88 | (3, 6, 8) | [(3, 12), (12, 11), (11, 10), (10, 9), (9, 5), (5, 6)] |
| 89 | (3, 6, 9) | [(3, 12), (12, 11), (11, 10), (10, 9), (9, 8), (8, 6)] |
| 90 | (3, 6, 10) | [(3, 12), (12, 11), (11, 10), (10, 16), (16, 8), (8, 6)] |
| 91 | (3, 7, 1) | [(3, 1), (1, 2), (2, 6), (6, 8), (8, 7)] |
| 92 | (3, 7, 2) | [(3, 4), (4, 5), (5, 6), (6, 8), (8, 7)] |
| 93 | (3, 7, 3) | [(3, 4), (4, 5), (5, 9), (9, 8), (8, 7)] |
| 94 | (3, 7, 4) | [(3, 4), (4, 11), (11, 10), (10, 9), (9, 8), (8, 7)] |
| 95 | (3, 7, 5) | [(3, 4), (4, 11), (11, 10), (10, 16), (16, 8), (8, 7)] |
| 96 | (3, 7, 6) | [(3, 4), (4, 11), (11, 10), (10, 16), (16, 18), (18, 7)] |
| 97 | (3, 7, 7) | [(3, 12), (12, 11), (11, 10), (10, 9), (9, 8), (8, 7)] |
| 98 | (3, 7, 8) | [(3, 12), (12, 11), (11, 10), (10, 16), (16, 8), (8, 7)] |
| 99 | (3, 7, 9) | [(3, 12), (12, 11), (11, 10), (10, 16), (16, 18), (18, 7)] |
| 100 | (3, 7, 10) | [(3, 1), (1, 2), (2, 6), (6, 5), (5, 9), (9, 8), (8, 7)] |
| 101 | (3, 18, 1) | [(3, 4), (4, 11), (11, 10), (10, 16), (16, 18)] |
| 102 | (3, 18, 2) | [(3, 12), (12, 11), (11, 10), (10, 16), (16, 18)] |
| 103 | (3, 18, 3) | [(3, 1), (1, 2), (2, 6), (6, 8), (8, 7), (7, 18)] |
| 104 | (3, 18, 4) | [(3, 1), (1, 2), (2, 6), (6, 8), (8, 16), (16, 18)] |
| 105 | (3, 18, 5) | [(3, 4), (4, 5), (5, 6), (6, 8), (8, 7), (7, 18)] |
| 106 | (3, 18, 6) | [(3, 4), (4, 5), (5, 6), (6, 8), (8, 16), (16, 18)] |
| 107 | (3, 18, 7) | [(3, 4), (4, 5), (5, 9), (9, 8), (8, 7), (7, 18)] |
| 108 | (3, 18, 8) | [(3, 4), (4, 5), (5, 9), (9, 8), (8, 16), (16, 18)] |
| 109 | (3, 18, 9) | [(3, 4), (4, 5), (5, 9), (9, 10), (10, 16), (16, 18)] |
| 110 | (3, 18, 10) | [(3, 4), (4, 11), (11, 10), (10, 17), (17, 16), (16, 18)] |
| 111 | (3, 20, 1) | [(3, 12), (12, 13), (13, 24), (24, 21), (21, 20)] |
| 112 | (3, 20, 2) | [(3, 4), (4, 11), (11, 10), (10, 15), (15, 19), (19, 20)] |
| 113 | (3, 20, 3) | [(3, 4), (4, 11), (11, 10), (10, 15), (15, 22), (22, 20)] |
| 114 | (3, 20, 4) | [(3, 4), (4, 11), (11, 10), (10, 16), (16, 18), (18, 20)] |
| 115 | (3, 20, 5) | [(3, 4), (4, 11), (11, 10), (10, 17), (17, 19), (19, 20)] |
| 116 | (3, 20, 6) | [(3, 4), (4, 11), (11, 14), (14, 15), (15, 19), (19, 20)] |
| 117 | (3, 20, 7) | [(3, 4), (4, 11), (11, 14), (14, 15), (15, 22), (22, 20)] |

Table J.1 (Continued)

| No | (origin, destination, route number) | Shortest paths |
|---|---|---|
| 118 | (3, 20, 8) | [(3, 4), (4, 11), (11, 14), (14, 23), (23, 22), (22, 20)] |
| 119 | (3, 20, 9) | [(3, 12), (12, 11), (11, 10), (10, 15), (15, 19), (19, 20)] |
| 120 | (3, 20, 10) | [(3, 12), (12, 11), (11, 10), (10, 15), (15, 22), (22, 20)] |
| 121 | (13, 6, 1) | [(13, 12), (12, 3), (3, 1), (1, 2), (2, 6)] |
| 122 | (13, 6, 2) | [(13, 12), (12, 3), (3, 4), (4, 5), (5, 6)] |
| 123 | (13, 6, 3) | [(13, 12), (12, 11), (11, 4), (4, 5), (5, 6)] |
| 124 | (13, 6, 4) | [(13, 12), (12, 11), (11, 10), (10, 9), (9, 5), (5, 6)] |
| 125 | (13, 6, 5) | [(13, 12), (12, 11), (11, 10), (10, 9), (9, 8), (8, 6)] |
| 126 | (13, 6, 6) | [(13, 12), (12, 11), (11, 10), (10, 16), (16, 8), (8, 6)] |
| 127 | (13, 6, 7) | [(13, 12), (12, 3), (3, 4), (4, 5), (5, 9), (9, 8), (8, 6)] |
| 128 | (13, 6, 8) | [(13, 12), (12, 11), (11, 4), (4, 3), (3, 1), (1, 2), (2, 6)] |
| 129 | (13, 6, 9) | [(13, 12), (12, 11), (11, 4), (4, 5), (5, 9), (9, 8), (8, 6)] |
| 130 | (13, 6, 10) | [(13, 12), (12, 11), (11, 10), (10, 17), (17, 16), (16, 8), (8, 6)] |
| 131 | (13, 7, 1) | [(13, 24), (24, 21), (21, 20), (20, 18), (18, 7)] |
| 132 | (13, 7, 2) | [(13, 12), (12, 11), (11, 10), (10, 9), (9, 8), (8, 7)] |
| 133 | (13, 7, 3) | [(13, 12), (12, 11), (11, 10), (10, 16), (16, 8), (8, 7)] |
| 134 | (13, 7, 4) | [(13, 12), (12, 11), (11, 10), (10, 16), (16, 18), (18, 7)] |
| 135 | (13, 7, 5) | [(13, 24), (24, 21), (21, 22), (22, 20), (20, 18), (18, 7)] |
| 136 | (13, 7, 6) | [(13, 24), (24, 23), (23, 22), (22, 20), (20, 18), (18, 7)] |
| 137 | (13, 7, 7) | [(13, 12), (12, 3), (3, 1), (1, 2), (2, 6), (6, 8), (8, 7)] |
| 138 | (13, 7, 8) | [(13, 12), (12, 3), (3, 4), (4, 5), (5, 6), (6, 8), (8, 7)] |
| 139 | (13, 7, 9) | [(13, 12), (12, 3), (3, 4), (4, 5), (5, 9), (9, 8), (8, 7)] |
| 140 | (13, 7, 10) | [(13, 12), (12, 11), (11, 4), (4, 5), (5, 6), (6, 8), (8, 7)] |
| 141 | (13, 18, 1) | [(13, 24), (24, 21), (21, 20), (20, 18)] |
| 142 | (13, 18, 2) | [(13, 12), (12, 11), (11, 10), (10, 16), (16, 18)] |
| 143 | (13, 18, 3) | [(13, 24), (24, 21), (21, 22), (22, 20), (20, 18)] |
| 144 | (13, 18, 4) | [(13, 24), (24, 23), (23, 22), (22, 20), (20, 18)] |
| 145 | (13, 18, 5) | [(13, 12), (12, 11), (11, 10), (10, 17), (17, 16), (16, 18)] |
| 146 | (13, 18, 6) | [(13, 24), (24, 23), (23, 22), (22, 21), (21, 20), (20, 18)] |
| 147 | (13, 18, 7) | [(13, 12), (12, 3), (3, 4), (4, 11), (11, 10), (10, 16), (16, 18)] |
| 148 | (13, 18, 8) | [(13, 12), (12, 11), (11, 10), (10, 9), (9, 8), (8, 7), (7, 18)] |
| 149 | (13, 18, 9) | [(13, 12), (12, 11), (11, 10), (10, 9), (9, 8), (8, 16), (16, 18)] |
| 150 | (13, 18, 10) | [(13, 12), (12, 11), (11, 10), (10, 15), (15, 19), (19, 20), (20, 18)] |
| 151 | (13, 20, 1) | [(13, 24), (24, 21), (21, 20)] |
| 152 | (13, 20, 2) | [(13, 24), (24, 21), (21, 22), (22, 20)] |
| 153 | (13, 20, 3) | [(13, 24), (24, 23), (23, 22), (22, 20)] |
| 154 | (13, 20, 4) | [(13, 24), (24, 23), (23, 22), (22, 21), (21, 20)] |
| 155 | (13, 20, 5) | [(13, 12), (12, 11), (11, 10), (10, 15), (15, 19), (19, 20)] |
| 156 | (13, 20, 6) | [(13, 12), (12, 11), (11, 10), (10, 15), (15, 22), (22, 20)] |
| 157 | (13, 20, 7) | [(13, 12), (12, 11), (11, 10), (10, 16), (16, 18), (18, 20)] |

Table J.1 (Continued)

| No | (origin, destination, | Shortest paths |
|---|---|---|
| | route number) | |
| 158 | (13, 20, 8) | [(13, 12), (12, 11), (11, 10), (10, 17), (17, 19), (19, 20)] |
| 159 | (13, 20, 9) | [(13, 12), (12, 11), (11, 14), (14, 15), (15, 19), (19, 20)] |
| 160 | (13, 20, 10) | [(13, 12), (12, 11), (11, 14), (14, 15), (15, 22), (22, 20)] |